\documentclass[traditabstract,longauth]{aa} 
\usepackage{amsmath}
\usepackage{graphicx}
\usepackage{txfonts}
\usepackage{natbib}
\usepackage{color}
\usepackage{multirow}
\usepackage{subfigure}
\usepackage{lscape}
\usepackage{latexsym}
\usepackage{changepage}		
\bibpunct{(}{)}{;}{a}{}{,} 

\usepackage[usenames,dvipsnames,svgnames,table]{xcolor}
\usepackage[breaklinks, colorlinks, citecolor=CornflowerBlue]{hyperref}
\usepackage{url}

\begin{document}
\definecolor{Red}{rgb}{1,0,0}
\authorrunning{Remco F.J. van der Burg, et al.}
   \title{The GOGREEN Survey: A deep stellar mass function of cluster galaxies at $1.0 < z < 1.4$ and the complex nature of satellite quenching}
   \titlerunning{The SMF of GOGREEN cluster galaxies at $1.0 < z < 1.4$}
   \author{Remco~F.~J.~van der Burg\inst{1}\thanks{\email{rvanderb@eso.org}}, Gregory Rudnick\inst{2}, Michael L. Balogh\inst{3,4}, Adam Muzzin\inst{5}, Chris Lidman\inst{6},\\ Lyndsay J. Old\inst{7,8}, Heath Shipley\inst{9}, David Gilbank\inst{10,11}, Sean McGee\inst{12}, 
Andrea Biviano\inst{13,14}, Pierluigi Cerulo\inst{15},\\
Jeffrey C. C. Chan\inst{16}, Michael Cooper\inst{17}, Gabriella De Lucia\inst{13}, 
Ricardo Demarco\inst{15}, Ben Forrest\inst{16}, Stephen Gwyn\inst{18}, Pascale Jablonka\inst{19,20}, Egidijus Kukstas\inst{21}, Danilo Marchesini\inst{22}, Julie Nantais\inst{23}, Allison Noble\inst{24,25}, \\ Irene Pintos-Castro\inst{8}, Bianca Poggianti\inst{26}, Andrew M. M. Reeves\inst{3,4}, Mauro Stefanon\inst{27}, \\Benedetta Vulcani\inst{26}, 
Kristi Webb\inst{3,4},  Gillian Wilson\inst{16}, Howard Yee\inst{8}, Dennis Zaritsky\inst{28} } 
	\institute{European Southern Observatory, Karl-Schwarzschild-Str. 2, 85748, Garching, Germany
	\and Department of Physics and Astronomy, The University of Kansas, 1251 Wescoe Hall Drive, Lawrence, KS 66045, USA
	\and Department of Physics and Astronomy, University of Waterloo, Waterloo, Ontario N2L 3G1, Canada
	\and Waterloo Centre for Astrophysics, University of Waterloo, Waterloo, ON, Canada N2L3G1
	\and Department of Physics and Astronomy, York University, 4700, Keele Street, Toronto, Ontario, ON MJ3 1P3, Canada
		\and The Research School of Astronomy and Astrophysics, Australian National University, ACT 2601, Australia
	\and European Space Agency (ESA), European Space Astronomy Centre, Villanueva de la Ca\~nada, E-28691 Madrid, Spain
	\and Department of Astronomy \& Astrophysics, University of Toronto, Toronto, Canada
		\and Department of Physics, McGill University, 3600 rue University, Montr\'eal, Qu\'ebec, H3P 1T3, Canada
	\and South African Astronomical Observatory, P.O. Box 9, Observatory 7935 Cape Town, South Africa
	\and Centre for Space Research, North-West University, Potchefstroom 2520 Cape Town, South Africa
	    \and School of Physics and Astronomy, University of Birmingham, Edgbaston, Birmingham B15 2TT, England
	\and INAF – Osservatorio Astronomico di Trieste, via G. B. Tiepolo 11, I-34143 Trieste, Italy
	\and IFPU – Institute for Fundamental Physics of the Universe, via Beirut 2, 34014 Trieste, Italy
	\and Departamento   de   Astronom\'ia,   Facultad   de   Ciencias   F\'isicas   y Matem\'aticas, Universidad de Concepci\'on, Concepci\'on, Chile
	\and Department of Physics and Astronomy, University of California, Riverside, 900 University Avenue, Riverside, CA 92521, USA
	\and Department of Physics and Astronomy, University of California, Irvine, 4129 Frederick Reines Hall, Irvine, CA 92697, USA
	\and NRC-Herzberg, 5071 West Saanich Road, Victoria, British Columbia V9E2E7, Canada
	\and Physics Institute, Laboratoire d’Astrophysique, Ecole Polytechnique F\'ed\'erale de Lausanne (EPFL), 1290 Sauverny, Switzerland 
       \and GEPI, Observatoire de Paris, Universit\'e PSL, CNRS, Place Jules Janssen, F-92190 Meudon, France
	\and Astrophysics Research Institute, Liverpool John Moores University, 146 Brownlow Hill, Liverpool L3 5RF, UK
		\and Department of Physics \& Astronomy, Tufts University, 574 Boston Avenue Suites 304, Medford, MA 02155, USA
	\and Departamento de Ciencias F\'isicas, Universidad Andres Bello, Fernandez Concha 700, Las Condes 7591538, Santiago, Regi\'on Metropolitana, Chile
	\and Arizona State University, School of Earth and Space Exploration, Tempe, AZ 871404, USA 
	\and MIT Kavli Institute for Astrophysics and Space Research, 70 Vassar St, Cambridge, MA 02109, USA
	\and INAF - Osservatorio astronomico di Padova, Vicolo Osservatorio 5, IT-35122 Padova, Italy
	\and Leiden Observatory, Leiden University, NL-2300 RA Leiden, The Netherlands
	\and Steward Observatory and Department of Astronomy, University of Arizona, Tucson, AZ 85719, USA
}

   \date{Submitted 17 February 2020; accepted 20 April 2020}

  \abstract{We study the stellar mass functions (SMFs) of star-forming and quiescent galaxies in 11 galaxy clusters at $1.0<z<1.4$, drawn from the Gemini Observations of Galaxies in Rich Early Environments (GOGREEN) survey. Based on more than 500 hours of Gemini/GMOS spectroscopy, and deep multi-band photometry taken with a range of observatories, we probe the SMFs down to a stellar mass limit of $10^{9.7}\,\rm{M_{\odot}}$ ($10^{9.5}\,\rm{M_{\odot}}$ for star-forming galaxies). At this early epoch, the fraction of quiescent galaxies is already highly elevated in the clusters compared to the field at the same redshift. The quenched fraction excess ($QFE$) represents the fraction of galaxies that would be star-forming in the field, but are quenched due to their environment. The $QFE$ is strongly mass dependent, and increases from $\sim$30\% at $M_{\star}=10^{9.7}\,\rm{M_{\odot}}$, to $\sim$80\% at $M_{\star}=10^{11.0}\,\rm{M_{\odot}}$. Nonetheless, the shapes of the SMFs of the two individual galaxy types, star-forming and quiescent galaxies, are identical between the clusters and the field - to high statistical precision. Yet, along with the different quiescent fractions is the total galaxy SMF environmentally dependent, with a relative deficit of low-mass galaxies in the clusters. These results are in stark contrast with findings in the local Universe, and thus require a substantially different quenching mode to operate at early times. We discuss these results in the light of several popular quenching models.}
   \maketitle
%

\section{Introduction}
Increasingly sophisticated statistical studies of the overall population of galaxies as a function of mass, cosmic time, and environment have provided a basic picture of the formation and evolution of galaxies \citep[e.g.][]{blantonmoustakas09,moster18,behroozi19}. While dark matter haloes continue to accrete material from their surrounding regions, some galaxies stop forming stars, or ``quench''. This leads to a distinct bimodality (in colour, star formation rate, morphology and other quantities) in the galaxy population \citep[e.g.][]{kauffmann03,baldry04,cassata08,wetzel12,taylor15}. The fraction of galaxies that are quenched depends strongly on their stellar mass, and also on their local environment \citep[e.g.][]{baldry06,peng10}. The physical drivers behind the overall process of quenching, and how these change with epoch and environment, are still poorly understood, and are thus a very active topic of extragalactic astronomy \citep[see][for a review]{somerville15}.

Quenching processes that are driven by internal mechanisms are referred to as mass- (or self-) quenching \citep{peng10}. In addition to this, there is an excess of quenched galaxies in over-dense environments \citep[environmental quenching, ][]{wetzel13}. This component can be quantified by local density, cluster-centric radius, or a general split between centrals and satellites. 
There is evidence that the quenching processes that are driven by mass and environment are largely separable, at least in the local ($z \lesssim 1 $) Universe \citep{baldry06,peng10,muzzin12,kovac14,guglielmo15,vdB18}. This is to say that there are no cross-terms; the effectiveness of environmental quenching does not depend on stellar mass, and the self/mass-quenching itself does not depend on the environment \citep[but for different interpretations, see][]{delucia12,contini20}. We note that the separability of these processes does not mean that they are physically unrelated processes; galaxies may quench due to shock heating of their cold gas component due to interactions with the hot (host) halo; this is generally referred to as ``halo quenching'', a process that may become efficient for both centrals and satellites in host-haloes $M_{\mathrm{halo}}\gtrsim 10^{12}\,\mathrm{M_{\odot}}$ \citep{dekel06,cattaneo08}.  

Since galaxy quenching takes place in an evolving density field, it is critical to constrain the physical mechanisms that lead to the ultimate quenching of galaxies as a function of cosmic epoch (i.e.~redshift). Indeed, gas accretion rates, consumption times, and the dynamical interactions between galaxies and their environments evolve rapidly with redshift. Furthermore, at fixed halo mass, dark matter haloes in over-dense environments form earlier \citep[assembly bias, cf.][]{zentner14,behroozi19}. Observationally, some studies find evidence that, at fixed stellar mass, galaxies with old stellar populations indeed favour regions of higher over-density \citep{cooper10b}, but see e.g.~\citet{lin16} for a null detection, albeit measured over much larger spatial scales. One would expect the efficiency of mass- and environmental quenching to evolve with redshift, and their separability might break down at an earlier epoch. There is growing evidence that the two modes of quenching are no longer acting fully independently at higher redshift \citep[$z\gtrsim 1$,][]{balogh16,kawinwanichakij17,papovich18,pintoscastro19}. 
In fact, the significant growth of central galaxies, and the build-up of intra-cluster light, may explain why the expected high abundance of low-mass quenched galaxies, as predicted by purely environmental quenching, is not observed in intermediate-redshift clusters \citep[cf.~][]{vdB18}. 

Exploring the effects of quenching in the early Universe ($z \gtrsim 1$) is a challenging task, especially when focussing on lower-mass ($M_{\star} \lesssim 10^{10}\,\mathrm{M_{\odot}}$) galaxies. Yet, studying the stellar-mass dependence of quenching over a wide range of masses is a good differentiator between models. In particular, at the lowest masses self-quenching is expected to be relatively ineffective so that environmental quenching processes become relatively more prominent \citep{geha12,peng12,wetzel12}. While typical contiguous surveys like COSMOS contain the necessary deep spectroscopy and photometry, they cover limited areas and contain only marginal over-densities. In contiguous surveys the density field is thus generally divided in density quartiles, or by density contrast $\delta$ \citep{cooper06,sobral11,davidzon16,darvish16,kawinwanichakij17,papovich18,lemaux19}. While such studies provide important constraints on the quenching of galaxies, the most extreme environmental conditions, which are found in galaxy clusters, are not explored. To be able to probe these environments, a sensible approach is to select galaxy clusters from wide-field surveys, and to specifically target cluster galaxies with extremely deep follow-up observations. We note that a hybrid approach was taken by the Observations of Redshift Evolution in Large-Scale Environments \citep[ORELSE][]{lubin09,tomczak17}, who quantify and study the large-scale environments around massive clusters at $0.6<z<1.0$.

We have recently completed the Gemini Observations of Galaxies in Rich Early ENvironments \citep[GOGREEN\footnote{http://gogreensurvey.ca/},][]{balogh17} survey, which is a deep spectroscopic (and multi-band photometric) survey of clusters and groups at $z \geq 1.0$. GOGREEN was designed to address some open questions related to galaxy quenching in highly over-dense environments at these epochs. Among the main science drivers of GOGREEN is a measurement of the relation between stellar mass and star formation in star-forming galaxies \citep[i.e.~the star forming main sequence, and how it depends on environment,][]{old20}. Furthermore, we wish to constrain quenching timescales by measuring the ages of quiescent galaxies in the clusters, and by comparing this to the co-eval (i.e.~at the same redshift) field (Webb et al., in prep.). Whereas earlier work based on the Gemini Cluster Astrophysics Spectroscopic Survey \citep[GCLASS,][]{muzzin12,vdB13} was restricted to stellar masses $M_{\star} \geq 10^{10.0}\,\mathrm{M_{\odot}}$, GOGREEN is designed to probe the galaxy population at lower masses, and to extend the sample to higher redshift. It will thus be more sensitive in the regime where model predictions are most discrepant \citep[e.g.][]{guo11,weinmann12,bahe17}. 

In this paper we measure the number density of galaxies as a function of stellar mass, i.e.~the stellar mass function (SMF) of galaxies in the GOGREEN clusters. 
Focussing primarily on the separate SMFs of star-forming and quiescent galaxies, this allows us to study what drives the quenching of star formation in galaxies at these early epochs ($z\gtrsim 1$). This work is an extension of local studies \citep[e.g.][]{balogh01,vulcani11,annunziatella14,annunziatella16}, who also used measurements of the galaxy SMF as a tool to understand galaxy transformations in cluster environments in terms of their morphology and star formation activity.

The structure of this paper is as follows. In Sect.~\ref{sec:sample} we describe the spectroscopic and photometric data set we use for the measurements. Most of the analysis is described in Sect.~\ref{sec:analysis}, and the results are presented in Sect.~\ref{sec:SMF}. To help interpret our findings, we discuss the measurements of the SMF in the context of several reference quenching models in Sect.~\ref{sec:discussion}. We conclude and summarise in Sect.~\ref{sec:summary}, and perform several robustness tests in the Appendices. 

All magnitudes we quote are in the Absolute Bolometric (AB) magnitude system, and we adopt $\Lambda$CDM cosmology with $\Omega_{\mathrm{m}}=0.3$, $\Omega_{\Lambda}=0.7$ and $\mathrm{H_0=70\, km\, s^{-1}\,  Mpc^{-1}}$. Uncertainties are given at the 1-$\sigma$ level, unless explicitly stated otherwise. Whenever results depend on the assumption of an Initial Mass Function (IMF), we will use the one from \citet{chabrier03}. We further explicitly note that, whenever we mention ``field'' in this work, we refer to an average/representative piece of Universe, which thus includes all environments.

\section{Cluster Sample \& Data}\label{sec:sample}
The cluster sample studied in this work is drawn from the GOGREEN survey \citep{balogh17}. The survey targets 21 systems that cover, by design, a range in redshift ($1.0<z<1.5$) and halo masses down to the group regime ($M_{200} \sim 5 \times 10^{13}\,\mathrm{M_{\odot}}$). 
GOGREEN targeted 12 clusters with $M_{200}\gtrsim 10^{14}\,\mathrm{M_{\odot}}$, 11 of which are studied in this paper\footnote{The full GOGREEN sample contains a twelfth cluster, SpARCS-1033, which is not included in the present work. This cluster is not yet covered by similarly deep multi-band photometry as the other 11. In the interest of studying a homogeneous sample, we have therefore not included it in this analysis.}. Three of those are clusters discovered by the South Pole Telescope (SPT) survey \citep{brodwin10, foley11, stalder13}. Eight others are lower-mass clusters taken from the Spitzer Adaptation of the Red-sequence Cluster Survey \citep{muzzin09,wilson09,demarco10}. For more details regarding the parent sample, we refer to \citet{balogh17} and the data release paper (Balogh et al., in prep.). Table~\ref{tab:dataoverview} in this paper gives an overview of the sample studied here. The following subsections summarise the photometric and spectroscopic components of our data set in turn.

\begin{table*}
\caption{Overview of the eleven GOGREEN clusters studied here.} 
\label{tab:dataoverview}
\begin{center}
\begin{tabular}{r l l l r r r r}
\hline
\hline
Name & RA$_{\mathrm{J2000}}^{\mathrm{BCG}}$ & Dec$_{\mathrm{J2000}}^{\mathrm{BCG}}$ & Redshift$^{\mathrm{a}}$ & $\lambda_{10.2, R<1000\,\mathrm{kpc}}^{\mathrm{b}}$ & IQ$^{\mathrm{c}}$ & $\mathrm{K_{s,lim}}^{\mathrm{d}}$  & $M_{\star,\mathrm{lim}}^{\mathrm{e}}$\\
&&&&&[$''$]&[mag$_{\mathrm{AB}}$]&[$\mathrm{M_{\odot}}$]\\
\hline
\texttt{SPTCL-0205} &02:05:48.19 & $-$58:28:49.0 & 1.320[106/31] &$41.1\pm7.7$&0.75 & 23.25 & 9.90\\
\texttt{SPTCL-0546} &05:46:33.67 & $-$53:45:40.6 & 1.067[156/70] &$94.1\pm10.6$&0.64 & 23.47 & 9.64\\
\texttt{SPTCL-2106} &21:06:04.59 & $-$58:44:27.9 & 1.132[$\,\,\,$95/56] &$108.6\pm11.2$&0.42 & 23.19 & 9.79\\
\texttt{SpARCS-0035} &00:35:49.68 & $-$43:12:23.8 & 1.335[326/33] &$45.2\pm7.9$&0.39 & 23.81 & 9.70\\
\texttt{SpARCS-0219} &02:19:43.56 & $-$05:31:29.6 & 1.325[338/12] &$22.2\pm6.3$&0.73 & 23.27 & 9.90\\
\texttt{SpARCS-0335} &03:35:03.56 & $-$29:28:55.8 & 1.368[133/32] &$32.4\pm7.1$&0.58 & 22.91 & 10.07\\
\texttt{SpARCS-1034} &10:34:49.47 & +58:18:33.1 & 1.386[$\,\,\,$84/24] &$20.8\pm6.2$&0.58 & 24.22 & 9.55\\
\texttt{SpARCS-1051} &10:51:11.23 & +58:18:02.7 & 1.035[199/48] &$13.5\pm5.7$&0.72 & 24.17 & 9.35\\
\texttt{SpARCS-1616} &16:16:41.32 & +55:45:12.4 & 1.156[243/70] &$49.4\pm8.1$&0.75 & 23.76 & 9.59\\
\texttt{SpARCS-1634} &16:34:37.00 & +40:21:49.3 & 1.177[191/69] &$35.8\pm7.2$&0.65 & 24.01 & 9.50\\
\texttt{SpARCS-1638} &16:38:51.64 & +40:38:42.9 & 1.196[192/68] &$18.7\pm5.9$&0.71 & 23.94 & 9.54\\
\hline
\end{tabular}
\end{center}
\begin{list}{}{}
\item[$^{\mathrm{a}}$] In brackets the number of spectroscopic redshifts overlapping with the region for which we have photometry, and the number of spectroscopic cluster members (here defined as being within 0.02 from the cluster mean redshift, which, depending on the cluster redshift, corresponds to a velocity cut of 2500-3000~km/s in the cluster rest-frame), respectively. We note that these cluster members are selected slightly differently from our other papers, where a selection was made in projected phase-space coordinates (cf.~Biviano et al. in prep.). This subtle difference is not relevant for the conclusions presented in this paper, and the approach followed here renders the membership selection more intuitive, when combined with photometric information.
\item[$^{\mathrm{b}}$] Richness, defined as the number of cluster members with $M_{\star} \geq 10^{10.2}\,\mathrm{M_{\odot}}$ that are found within a circular aperture with $R<1000\,\mathrm{kpc}$. This parameter is used to scale galaxy counts in the SMF in low-mass bins where not every cluster contributes to the measurement of the SMF due to incompleteness.
\item[$^{\mathrm{c}}$] FWHM of the PSF measured in the detection image ($\mathrm{K_s}$-band).
\item[$^{\mathrm{d}}$] Faintest magnitude at which 80\% of injected sources are still recovered. More details are given in Sect.~\ref{sec:completeness}.
\item[$^{\mathrm{e}}$] Stellar mass limit based on a relatively old stellar population, as described in Sect.~\ref{sec:completeness}. This is the stellar mass limit we adopt for the quiescent population. For star-forming galaxies, which are brighter for their stellar mass, we expect to probe 0.2 \texttt{dex} below this limit.

\end{list}
\end{table*}

\subsection{Cluster spectroscopy}
Our deep Gemini/GMOS spectroscopy forms the backbone of this analysis. The main data set is taken with a $\sim$400-hour Gemini Large Program (PI=Balogh, GS LP-1 and GN LP-4). Five of the clusters were also part of GCLASS, which resulted in additional spectroscopic coverage ($\sim$100 hours) for the brighter galaxies of these clusters.

The spectroscopic target galaxies for GOGREEN were selected based on $\mathrm{[3.6]\,\mu m}$ imaging obtained from different \textit{Spitzer}/IRAC programs \citep[primarily SERVS and SWIRE;][]{lonsdale03,mauduit12}, in combination with deep Gemini GMOS $z$-band pre-imaging which we obtained as part of the survey. 

\citet{balogh17} have identified a region in a $z-\mathrm{[3.6]}$ versus $z$ colour-magnitude diagram, where the purity and completeness of selecting galaxies in the redshift range $1.0<z<1.5$ is high. Targeting these with the highest priority, the observing strategy chosen by GOGREEN is such that the fainter galaxies appear in multiple slit masks, resulting in integration times of up to 15 hours. Since individual masks are exposed for 3 hours, slits on brighter targets can change more frequently. This ensures a high spectroscopic completeness (and a high success rate in measuring reliable redshifts) over a large baseline of magnitudes (or stellar masses). The procedure is laid out in more detail in Sect.~2.4 of \citet{balogh17}.

To the GOGREEN and GCLASS spectroscopy we add an existing body of literature redshifts from different sources. SPT has taken spectra to confirm and characterise their three clusters \citep{brodwin10,foley11,stalder13}. One of those clusters, SPTCL-0546, is also part of the Atacama Cosmology Telescope (ACT) survey, and we have included redshifts measured by \citet{sifon13}. The PRIsm MUlti-object Survey \citep[PRIMUS,][]{coil11b,cool13} overlaps with two of our clusters, one of which is also covered by the VIMOS Public Extragalactic Redshift Survey \citep[VIPERS,][]{scodeggio18}. One cluster, \texttt{SpARCS-0335}, was also studied in \citet{nantais16}, and we use the redshifts measured with VLT/FORS2 from their work. Furthermore, seven clusters are covered in DR14 of the SDSS \citep{sdssDR14}. We note that not all these literature sources provide deep enough spectroscopy to allow for the identification of additional cluster members, but they nonetheless provide redshifts over a wider baseline, such that we can calibrate and test our photometric redshifts.

\subsection{Cluster photometric data}
\begin{table*}
\caption{Illustration of the photometric data set used in this work. The reported depths are median 5-$\sigma$ limits measured on the PSF-homogenized stacked images in circular apertures with a diameter of 2$''$. The values listed are after correction for Galactic dust extinction, so are indicative of the galaxy population we study. The instruments and filters used for the different clusters are indicated. For IRAC we measure fluxes in apertures with a diameter of 3$''$ and convert them back to 2$''$ by using the detection band, convolved to both PSF sizes, as described in Sect.~\ref{sec:objectdetection}. For reference, we list the median 5-$\sigma$ depths of 13 stacks in the DR1 COSMOS/UltraVISTA catalogues \citep[these are measured in 2\farcs1 apertures][]{muzzin13a}. That subset of filters of the full COSMOS/UltraVISTA data set is used to provide a statistical background correction in this work, and thus to provide a verification of our fiducial method to measure the cluster SMF (cf.~Sect.~\ref{sec:membership}).}
\label{tab:photometry}
\begin{adjustwidth}{-0.6cm}{}
\begin{tabular}{l l l l l l l l l l l l l l}
\hline
\hline
Name &$u/U$&$B/g$&$V$&$r/R$&$i/I$&$z/Z$&$Y/J_1$&$J$&$K_\mathrm{s}$&[3.6] $\mu$m&[4.5] $\mu$m&[5.8] $\mu$m&[8.0] $\mu$m\\
\hline
\texttt{SPTCL-0205}
&26.2$^\mathrm{b}$&26.7$^\mathrm{b}$&25.6$^\mathrm{b}$&25.9$^\mathrm{b}$&25.4$^\mathrm{b}$&24.2$^\mathrm{b}$&24.2$^\mathrm{h}$&23.9$^\mathrm{h}$&24.0$^\mathrm{h}$&23.7$^\mathrm{j}$&23.2$^\mathrm{j}$&$-$&$-$\\
\texttt{SPTCL-0546}
&25.3$^\mathrm{b}$&26.1$^\mathrm{b}$&25.3$^\mathrm{b}$&25.6$^\mathrm{b}$&25.0$^\mathrm{b}$&23.8$^\mathrm{b}$&24.1$^\mathrm{h}$&23.9$^\mathrm{h}$&23.9$^\mathrm{h}$&24.0$^\mathrm{j}$&23.8$^\mathrm{j}$&$-$&$-$\\
\texttt{SPTCL-2106}
&26.0$^\mathrm{b}$&26.3$^\mathrm{b}$&25.9$^\mathrm{b}$&25.8$^\mathrm{b}$&25.3$^\mathrm{b}$&24.6$^\mathrm{b}$&24.4$^\mathrm{h}$&24.1$^\mathrm{h}$&23.6$^\mathrm{g}$&23.7$^\mathrm{j}$&23.0$^\mathrm{j}$&$-$&$-$\\
\texttt{SpARCS-0219}
&25.8$^\mathrm{b}$&26.0$^\mathrm{b}$&25.3$^\mathrm{b}$&25.5$^\mathrm{b}$&25.2$^\mathrm{b}$&24.1$^\mathrm{b}$&24.4$^\mathrm{h}$&24.3$^\mathrm{h}$&24.0$^\mathrm{h}$&24.0$^\mathrm{j}$&23.8$^\mathrm{j}$&21.4$^\mathrm{j}$&21.4$^\mathrm{j}$\\
\texttt{SpARCS-0335}
&26.3$^\mathrm{b}$&26.4$^\mathrm{b}$&25.9$^\mathrm{b}$&26.3$^\mathrm{b}$&25.5$^\mathrm{b}$&24.6$^\mathrm{b}$&25.2$^\mathrm{g}$&24.3$^\mathrm{h}$&23.7$^\mathrm{h}$&24.4$^\mathrm{j}$&24.3$^\mathrm{j}$&21.6$^\mathrm{j}$&21.6$^\mathrm{j}$\\
\texttt{SpARCS-0035}
&25.9$^\mathrm{b}$&26.4$^\mathrm{b}$&25.8$^\mathrm{b}$&26.0$^\mathrm{b}$&25.5$^\mathrm{b}$&25.5$^\mathrm{d}$&24.2$^\mathrm{h}$&24.9$^\mathrm{g}$&24.2$^\mathrm{g}$&24.6$^\mathrm{j}$&24.5$^\mathrm{j}$&22.8$^\mathrm{j}$&22.6$^\mathrm{j}$\\
\texttt{SpARCS-1034}
&$-$&26.0$^\mathrm{c}$&$-$&26.1$^\mathrm{c}$&25.5$^\mathrm{c}$&25.4$^\mathrm{e}$&25.1$^\mathrm{e}$&24.5$^\mathrm{i}$&24.0$^\mathrm{i}$&22.7$^\mathrm{j}$&22.4$^\mathrm{j}$&19.9$^\mathrm{j}$&19.7$^\mathrm{j}$\\
\texttt{SpARCS-1051}
&26.3$^\mathrm{a}$&26.1$^\mathrm{c}$&$-$&26.1$^\mathrm{c}$&25.6$^\mathrm{c}$&25.4$^\mathrm{e}$&25.0$^\mathrm{e}$&24.5$^\mathrm{i}$&24.1$^\mathrm{i}$&22.6$^\mathrm{j}$&22.5$^\mathrm{j}$&19.7$^\mathrm{j}$&19.6$^\mathrm{j}$\\
\texttt{SpARCS-1616}
&25.9$^\mathrm{a}$&26.2$^\mathrm{c}$&$-$&26.1$^\mathrm{c}$&25.7$^\mathrm{c}$&25.6$^\mathrm{e}$&24.7$^\mathrm{e}$&24.2$^\mathrm{i}$&23.8$^\mathrm{i}$&22.7$^\mathrm{j}$&22.6$^\mathrm{j}$&21.2$^\mathrm{j}$&21.3$^\mathrm{j}$\\
\texttt{SpARCS-1634}
&25.9$^\mathrm{a}$&26.4$^\mathrm{c}$&$-$&26.2$^\mathrm{c}$&25.8$^\mathrm{c}$&25.0$^\mathrm{f}$&$-$&24.2$^\mathrm{i}$&23.8$^\mathrm{i}$&23.0$^\mathrm{j}$&22.8$^\mathrm{j}$&21.3$^\mathrm{j}$&21.3$^\mathrm{j}$\\
\texttt{SpARCS-1638}
&26.1$^\mathrm{a}$&26.4$^\mathrm{c}$&$-$&26.2$^\mathrm{c}$&25.6$^\mathrm{c}$&25.3$^\mathrm{f}$&24.2$^\mathrm{c}$&24.1$^\mathrm{i}$&23.6$^\mathrm{i}$&22.8$^\mathrm{j}$&22.5$^\mathrm{j}$&21.3$^\mathrm{j}$&21.4$^\mathrm{j}$\\
\hline
COSMOS/  & \multirow{2}{*}{$26.8^\mathrm{a}$}& \multirow{2}{*}{$26.9^\mathrm{c}$}& \multirow{2}{*}{$26.4^\mathrm{c}$}& \multirow{2}{*}{$26.4^\mathrm{c}$}& \multirow{2}{*}{$26.0^\mathrm{c}$}& \multirow{2}{*}{$25.2^\mathrm{c}$}& \multirow{2}{*}{$24.5^\mathrm{k}$}& \multirow{2}{*}{$24.3^\mathrm{k}$}& \multirow{2}{*}{$23.8^\mathrm{k}$}& \multirow{2}{*}{$23.9^\mathrm{j}$}& \multirow{2}{*}{$23.6^\mathrm{j}$}& \multirow{2}{*}{$21.7^\mathrm{j}$}& \multirow{2}{*}{$21.7^\mathrm{j}$}\\
UltraVISTA \\
\hline
\end{tabular}
\end{adjustwidth}
\hspace{1cm} $^{\mathrm{a}}$ CFHT/MegaCam, $^{\mathrm{b}}$ VLT/VIMOS, $^{\mathrm{c}}$ Subaru/SuprimeCam, $^{\mathrm{d}}$ Blanco/DECam, $^{\mathrm{e}}$ Subaru/HSC, $^{\mathrm{f}}$ Gemini/GMOS, \hfill 

\hspace{1cm} $^{\mathrm{g}}$ VLT/HAWKI, $^{\mathrm{h}}$ Magellan/FourStar, $^{\mathrm{i}}$ CFHT/WIRCam, $^{\mathrm{j}}$ \textit{Spitzer}/IRAC, $^{\mathrm{k}}$ VISTA/VIRCAM
\end{table*}

The multiband photometry that we have obtained for the GOGREEN clusters serves several important purposes. Whereas the GMOS spectroscopy only covers a wavelength range from 6,400$\AA$ up to about 10,200$\AA$, using multi-band photometry we can characterise the galaxy SEDs more accurately, and provide further constraints on their star-forming properties and stellar masses.  In particular, based on photometry taken at rest-frame wavelengths ranging from the UV to $J$-band, we can characterise galaxies in terms of their general type (quiescent versus star forming) and dust extinction. Furthermore, even a spectroscopic program like GOGREEN is, due to practical limitations, not complete - it has not targeted all cluster members. Based on accurate and precise photometric redshifts we characterise the parent galaxy population from which the spectroscopic targets were selected. A combination of this information is essential if we are to make a measurement of the entire cluster galaxy population (as required to measure the SMF). 

The cluster sample covers a range in declinations between the North and South. Together with the wide coverage in wavelength we are aiming for, this required us to utilise multiple telescope sites and instruments. Table~\ref{tab:photometry} lists all telescopes and instruments that form the basis of the current photometric analysis. The exposure times, and associated depths, of our photometry were tailored to allow for an unbiased detection of galaxies with stellar masses down to the $10^{9.5}\,\lesssim\,M_{\star}/\mathrm{M_{\odot}}\lesssim\,10^{10.0}$ range at the redshifts of the GOGREEN clusters, and to characterise the detected sources by means of their broad-band SEDs. These steps are described in detail in Sect.~\ref{sec:analysis}. 

All photometric data sets undergo basic reduction steps such as flat-fielding, cosmic-ray rejection, astrometric registering and background subtraction. Especially for the Near-IR data, a proper data reduction relies on a dithered set of exposures to perform the sky background subtraction. The astrometric registering is done with \texttt{SCAMP} \citep{scamp}, using the USNO-B1 catalogue \citep{monet03}. Astrometry is aligned well within 0.10$''$ between filters, ensuring reliable colour measurements.

We mask regions of the images that are not suitable for our analysis. First, we mask bright stars, their diffraction spikes and reflective haloes, as well as artefacts in any photometric band. We also, conservatively, require that photometry in all bands listed in Table~\ref{tab:photometry} is available at any sky position considered in this work. This ensures a study with a similar data set per cluster. Since data are taken with a range of different telescopes and instruments, the area considered for this study ranges from $\sim 5\times 5'$ to $\sim 10\times 10'$. In the most restricted analysis, where we rely on the Gemini/GMOS $z$-band pre-imaging for our photometric analysis, we still probe the galaxy population to radial distances of $\sim$1500~kpc from the cluster centres, well beyond the cluster virial radius or $R_{200}$ (Biviano et al., in prep.).

\subsection{Cluster centres - Brightest Cluster Galaxies}
The analysis presented in this paper is performed with respect to the cluster centres defined as the positions of the Brightest Cluster Galaxies (BCGs). The identification of BCGs in these clusters is not always straightforward, as some clusters at high-$z$ have BCGs that are significantly less dominant in terms of brightness compared to the overall galaxy population \citep{lidman12}, and one given galaxy is not always the brightest one in every photometric band. In this work we define the BCG as the most massive galaxy with a photometric redshift consistent with the cluster mean redshift, and projected within 500 kpc from the main galaxy over-density. In general, these candidates correspond to the galaxies that are brightest in the redder photometric bands. In some cases, notably SPTCL-0546, our \textit{HST} F160W photometry (PI=Wilson, PID=15294; Chan et al., in prep.) helped to separate a dense clump of neighbouring galaxies that were blended in the ground-based photometry, to revise our identification of the BCG. 

Five of our clusters overlap with the BCG sample studied in \citet{lidman12}. In four cases we identify the same BCGs as in that work, but for SpARCS-1634 we note that our \textit{HST} F160W photometry identifies the BCG candidate from \citet{lidman12} as a major merger rather than a single massive galaxy. The coordinates of our final sample of BCG candidates are in Table~\ref{tab:dataoverview}. In all but two cases, these candidate BCGs were spectroscopically targeted, and thus securely confirmed to be part of the cluster. The exceptions are SPTCL-2106 and SpARCS-0219, for which we have to rely on photometric information. We note that in this work, where we study the cluster galaxy SMF, the results are not strongly affected by how we define the cluster centres\footnote{We note that our study does not treat BCGs differently from other/satellite galaxies, and they are included in the measurement of the SMF.}. 

Appendix~\ref{app:colourimages} presents colour images for each cluster, based on three photometric bands. The cut-outs are centred on the BCG locations, and spectroscopic targets are marked (cluster members in green).

\section{Analysis}\label{sec:analysis}
\subsection{Object detection and photometry}\label{sec:objectdetection}
We perform object detection in the original, un-convolved, $K_\mathrm{s}$-band stacks by running \texttt{SExtractor} \citep{bertinarnouts96} with the requirement that sources have at least 5 adjacent pixels that are $>$1.5$\sigma$ above the local background RMS.  

To perform aperture photometry on the same intrinsic part of each source, we convolve each individual stack with a kernel created with \texttt{PSFEx} \citep{psfex} to bring them to a common (Moffat-shaped) point spread function (PSF) for each cluster field. Aperture photometry is measured on these homogenised stacks, using circular apertures with a diameter of 2$''$. 

A standard approach would be to convolve all images to match the image with the worst PSF, which is the \textit{Spitzer}/IRAC data. However, to benefit from the superior spatial information from the ground-based imaging, we incorporate aperture fluxes measured in IRAC following the approach that was laid out in \citet{vdB13}, and introduced by \citet{quadri07}. In addition to the IRAC channels, we convolve only the $K_\mathrm{s}$-band stack to the largest FWHM PSF (Moffat with FWHM 2.0$''$ or 2.5$''$, depending on whether a cluster has been observed in IRAC [5.8] and [8.0]$\mu$m, cf.~Table~\ref{tab:photometry}). Then, all IRAC fluxes are measured within apertures that have a diameter of 3$''$, so that the IRAC PSF size is better matched than with the original, smaller, apertures. The flux we use in the SED fitting, which is included in the photometric catalogues, is defined as:
\begin{equation}
\mathrm{IRAC_{cat}} = \mathrm{IRAC_{largePSF,3'' app}} \times \frac{K\mathrm{s_{smallPSF,2'' app}}}{K\mathrm{s_{largePSF,3'' app}}}
\end{equation}
This approach largely removes source confusion and blending as it accounts for the contribution of neighbouring sources whose fluxes leak into the IRAC aperture, under the assumption that the $K_\mathrm{s}$-IRAC colours are similar for the studied source as for the contaminant. 

In order to perform aperture photometry on stacks other than the IRAC imaging, we consider the stacks with the worst image quality per cluster, IQ$_{\mathrm{max,cl}}$, which have FWHMs ranging from 0\farcs83 to 1\farcs28. The \texttt{PSFEx} kernels convolve each stack to a PSF with a Moffat-$\beta$ parameter of 2.5 and a FWHM of 1.1$\times$IQ$_{\mathrm{max,cl}}$+0.05. These choices ensure that the target PSF has sufficiently broad wings that no deconvolution is required.

Since our analysis is focussed on faint galaxies, uncertainties on aperture flux measurements are dominated by fluctuations in the background. To estimate this noise component, we randomly place apertures on sky positions that do not overlap with sources that are detected in the $K_\mathrm{s}$-band. The resulting fluxes approximately follow a Gaussian distribution centred around 0. The RMS, which depends on the local image depth, defines the flux uncertainty. The depths quoted in Table~\ref{tab:photometry} correspond to the median depth of the unmasked area, measured on PSF-homogenised images. 

Relative flux calibration (i.e.~for measuring colours) is done based on the universal properties of the stellar locus \citep{slr,kelly14}. We consider the wavelength response of each photometric observation independently. Effective wavelength response curves are obtained by considering the throughput of the telescopes, the detector response, the used filters\footnote{cf.~http://svo2.cab.inta-csic.es/theory/fps/ for a large compilation of filter throughput curves.} and atmosphere transmission models. We obtain a reference stellar locus for each combination of filters by integrating stellar libraries from \citet{pickles98}, for which flux measurements are taken in the Near-IR by \citet{ivanov04}. In addition, we also consider the library used by \citet{kelly14} and integrate all these stellar spectra through the effective response curves. Our photometry is then calibrated by applying offsets to the instrumental magnitudes, so that stellar colours match the reference locus. 
We note that Galactic dust extinction is negligible in the fields we study. 

These calibration steps lead to colour calibrations with typically $\sim 0.01-0.03$ mag uncertainties. We chose the anchor point for the absolute flux calibration to be the 2MASS all-sky catalogue of point sources \citep{2MASS}, to which we match our total $J$- and $K_{\mathrm{s}}$-band instrumental magnitudes. Those total instrumental magnitudes are measured with \texttt{SExtractor} in Kron-like apertures (option \texttt{MAG\_AUTO}). This allows a flux measurement that is only slightly lower than the total/intrinsic value. We make a $\sim$0.02-0.10 magnitude correction based on source simulations, as detailed in Sect.~\ref{sec:completeness} and Appendix \ref{app:eddington}.

\subsection{Photometric redshifts}
We estimate photometric redshifts for our sources using the template-fitting code \texttt{EAZY} \citep[Version May 2015;][]{brammer08}. The basic EAZY templates are used, which are based on the PEGASE model library \citep{fioc97}, in addition to a red galaxy template taken from \citet{maraston05}. In the following we refer to $z_{\mathrm{phot}}$ as the peak of the posterior probability distribution of the redshift estimated with \texttt{EAZY}. To quantify the quality of the measured photometric redshifts, we define a relative scatter  $ \Delta z =\frac{z_{\mathrm{phot}} -z_{\mathrm{spec}}}{1+z_{ \mathrm{spec}}} $ for each object that has a reliable spectroscopic redshift $z_{\mathrm{spec}}$. 

Initially, this process results in 4.7\% outliers, defined here as objects for which $|\Delta z |> 0.15$. For the remaining galaxies, we measure a bias of -0.03 ($z_{\mathrm{phot}}$ values are slightly too low compared to $z_{\mathrm{spec}}$), and a scatter around the mean of 0.043. 

We find a subtle, but significant, residual trend between the estimated $z_{\mathrm{phot}}$ and $z_{\mathrm{spec}}$, which suggests that the initial $z_{\mathrm{phot}}$ estimates are not optimal. This may be due to small residuals in the photometric calibration, for example because the typical atmosphere models that are included in the filter throughputs are not fully representative of the atmospheric conditions at the time of the observations. 
Rather than re-training the photometric calibration based on these offsets, we find that these residuals are well described by the quadratic functions $z_{\mathrm{phot}}=1.12\times z_{\mathrm{EAZY}}-0.03\times z_{\mathrm{EAZY}}^2$ for the Southern clusters, and $z_{\mathrm{phot}}=z_{\mathrm{EAZY}}+0.05$ for the Northern clusters. After correcting for these residuals, we are left with $\sim4.1 \%$ outliers, a mean $\Delta z$ of 0 (by construction there is no bias after correction), and a scatter around this mean of $\sigma_z$=0.048. 

\begin{figure*}
\resizebox{\hsize}{!}{\includegraphics{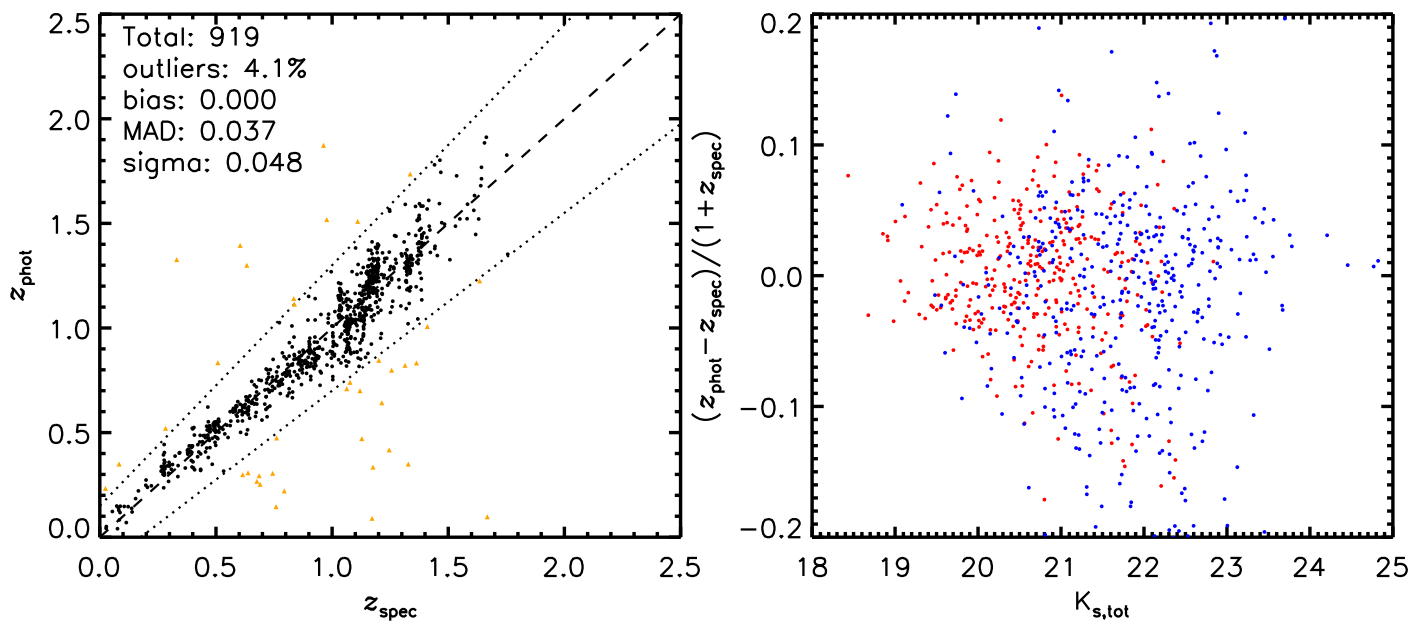}}
\caption{\textit{Left:} Photometric versus spectroscopic redshifts for all galaxies in the 11 cluster fields. Outliers, defined as objects for which $|\Delta z| > 0.15$, are marked in orange. The outlier fraction is $4.1\%$, the scatter of the remaining objects is $\sigma_z = 0.048$.  \textit{Right:} $\Delta z$ as a function of $K_\mathrm{s}$-band magnitude, for sources with $ 1.0<z_{\mathrm{spec}}<1.5$. Quiescent and star-forming galaxies (separated according to the criteria given in Sect.~\ref{sec:rfcolours}) are marked in red and blue, respectively.}
\label{fig:speczphotz}
\end{figure*}

Figure \ref{fig:speczphotz} compares the spectroscopic and photometric redshifts, after the correction. We note that these statistics are measured for galaxies more massive than $10^{10}\,\mathrm{M_{\odot}}$, even though the correction was applied to all galaxies. If we instead consider those more massive than $10^{9.5}\,\mathrm{M_{\odot}}$, the outlier fraction increases to 7.0\% and the scatter increases slightly to 0.045 in $\Delta z$. 

The $z_{\mathrm{phot}}$ estimator does not straightforwardly identify stellar objects among the galaxy population. Rather, we make a distinction between stars and galaxies based on their different broad-band colours. We apply a similar selection as previous studies \citep[e.g.][]{whitaker11,vdB18}, to select the sample of galaxies: 
\begin{align}
 J-K_\mathrm{s} &>0.18\cdot(u- J )-0.60 \,\cup \\
 J-K_\mathrm{s} &>0.08\cdot(u- J )-0.30\,.
\end{align}  
Where there is no $u$-band data available, we use the $g$-band instead, and assume a typical colour ($u-g$)=0.7 to shift the selection region:
\begin{align}
 J-K_\mathrm{s} &>0.18\cdot(g- J )-0.47 \,\cup \\
 J-K_\mathrm{s} &>0.08\cdot(g- J )-0.24\,.
\end{align}  
We verify these selection criteria by considering the measured colour distributions, which indeed show a clear separation between the cloud of galaxies and the stellar locus. 
We also considered a separation between stars and galaxies based on their spatial extent compared to the size of the PSF. We find that this provides a similar selection for brighter sources, whereas the broad-band colours outperform a morphological selection at the faint end of the source distribution.

\subsection{Stellar masses and galaxy types}\label{sec:rfcolours}
Stellar masses are inferred for each galaxy based on the total $K_\mathrm{s}$-band instrumental magnitude, and using the SED-fitting code FAST \citep{kriek09}, which uses stellar population synthesis models from \citet{bc03}. We assume a \citet{chabrier03} IMF, solar metallicity, and the dust law from \citet{calzetti00}. Following the UltraVISTA reference sample, we parameterise the star formation history as $\mathrm{SFR} \propto e^{-t/\tau} $, where the timescale $\tau$ ranges between 10 Myr and 10 Gyr, and the age (onset of star formation) is left as another free parameter. Star-formation histories that are parametrised in this way may underestimate the stellar mass by $\sim$0.2 \texttt{dex} compared to when star-formation histories are estimated in bins \citep[][Webb et al., in prep.]{leja19}. However, since our goal is to perform a consistent relative comparison with the UltraVISTA field survey, we use the same parameterisation as used there \citep{muzzin13a}.

\begin{table*}
\caption{Data points of the SMFs measured in this work. Error bars for the cluster SMFs are based on bootstrap resamplings, where clusters are drawn with replacement as detailed in the text. The quiescent galaxies in the cluster environment are only reliably detected and characterised at stellar masses $M_{\star} \geq 10^{9.7}\,\mathrm{M_{\odot}}$.}
\label{tab:datapoints}
\begin{adjustwidth}{-0.8cm}{}
\begin{tabular}{l || l l l | l l l | r r r}
\hline
\hline
&\multicolumn{3}{c}{Cluster$<$1000 kpc}&\multicolumn{3}{c}{Cluster$<$500 kpc}&\multicolumn{3}{c}{Field} \\
&\multicolumn{3}{c}{$\Phi\, [$cluster$^{-1}$ dex$^{-1}]$}&\multicolumn{3}{c}{$\Phi\, [$cluster$^{-1}$ dex$^{-1}]$}&\multicolumn{3}{c}{$\Phi\, [10^{-5}$ Mpc$^{-3}$ dex$^{-1}]$}\\
$\mathrm{log[M_{\star} /\mathrm{M_{\odot}}]}$ & All & Quiescent & Star-forming & All & Quiescent & Star-forming & All & Quiescent & Star-forming \\
\hline
\,\,\,9.55&\,\,\,\,\,\,\,\,\,-\,\,\,\,\,\,\,\,\,&\,\,\,\,\,\,\,\,\,-\,\,\,\,\,\,\,\,\,&$96.5^{+95.9}_{-17.6}$&\,\,\,\,\,\,\,\,\,-\,\,\,\,\,\,\,\,\,&\,\,\,\,\,\,\,\,\,-\,\,\,\,\,\,\,\,\,&$60.0^{+72.0}_{-15.9}$&$478.4\pm15.1$&$32.9\pm4.1$&$445.4\pm14.5$\\
\,\,\,9.65&\,\,\,\,\,\,\,\,\,-\,\,\,\,\,\,\,\,\,&\,\,\,\,\,\,\,\,\,-\,\,\,\,\,\,\,\,\,&$47.4^{+14.2}_{-2.1}$&\,\,\,\,\,\,\,\,\,-\,\,\,\,\,\,\,\,\,&\,\,\,\,\,\,\,\,\,-\,\,\,\,\,\,\,\,\,&$27.9^{+18.8}_{-6.4}$&$452.6\pm10.1$&$32.8\pm2.8$&$419.8\pm9.7$\\
\,\,\,9.75&$34.9^{+15.2}_{-4.8}$&\,\,\,$7.7^{+14.1}_{-3.9}$&$27.2^{+5.7}_{-3.5}$&$19.4^{+17.5}_{-5.7}$&\,\,\,$6.7^{+13.4}_{-4.2}$&$12.7^{+5.0}_{-2.9}$&$369.7\pm8.1$&$32.1\pm2.4$&$337.6\pm7.7$\\
\,\,\,9.85&$59.3^{+13.5}_{-6.1}$&$23.5^{+6.7}_{-2.9}$&$35.8^{+9.3}_{-4.6}$&$23.9^{+8.5}_{-5.1}$&$12.2^{+5.6}_{-4.5}$&$11.8^{+3.0}_{-2.1}$&$379.3\pm8.2$&$41.3\pm2.7$&$338.0\pm7.7$\\
\,\,\,9.95&$63.2^{+31.3}_{-5.3}$&$29.9^{+18.1}_{-4.3}$&$33.3^{+6.3}_{-3.5}$&$27.9^{+13.9}_{-4.7}$&$16.2^{+10.8}_{-2.8}$&$11.8^{+3.4}_{-3.5}$&$326.2\pm7.6$&$53.3\pm3.1$&$272.9\pm6.9$\\
10.05&$67.5^{+17.5}_{-1.7}$&$37.2^{+7.5}_{-4.2}$&$30.3^{+11.5}_{-2.1}$&$37.5^{+10.5}_{-2.3}$&$26.1^{+8.8}_{-3.7}$&$11.4^{+6.5}_{-2.6}$&$282.2\pm7.0$&$54.5\pm3.1$&$227.7\pm6.3$\\
10.15&$63.2^{+16.4}_{-7.3}$&$30.7^{+8.2}_{-4.1}$&$32.5^{+13.2}_{-6.4}$&$36.6^{+9.3}_{-5.6}$&$19.7^{+4.4}_{-2.6}$&$16.9^{+5.5}_{-5.2}$&$298.4\pm7.2$&$68.0\pm3.5$&$230.4\pm6.3$\\
10.25&$70.3^{+19.5}_{-8.9}$&$40.4^{+14.1}_{-8.3}$&$30.0^{+8.3}_{-3.5}$&$35.6^{+11.3}_{-6.2}$&$23.1^{+8.3}_{-6.8}$&$12.5^{+5.5}_{-3.6}$&$249.4\pm6.6$&$73.3\pm3.6$&$176.0\pm5.5$\\
10.35&$67.6^{+9.2}_{-4.9}$&$41.0^{+6.9}_{-4.7}$&$26.6^{+6.3}_{-4.9}$&$34.4^{+5.7}_{-3.6}$&$24.4^{+3.9}_{-4.1}$&$10.0^{+3.6}_{-2.6}$&$267.5\pm6.8$&$86.2\pm3.9$&$181.3\pm5.6$\\
10.45&$73.8^{+13.3}_{-8.4}$&$47.3^{+9.0}_{-8.3}$&$26.5^{+10.4}_{-4.5}$&$39.6^{+11.3}_{-4.0}$&$27.2^{+5.1}_{-4.9}$&$12.4^{+9.1}_{-4.9}$&$222.8\pm6.2$&$80.8\pm3.7$&$142.0\pm4.9$\\
10.55&$60.3^{+8.9}_{-10.4}$&$39.1^{+6.0}_{-7.8}$&$21.2^{+4.7}_{-5.7}$&$29.3^{+4.8}_{-2.5}$&$20.5^{+4.8}_{-2.9}$&\,\,\,$8.8^{+4.4}_{-3.4}$&$203.4\pm5.9$&$83.4\pm3.8$&$119.9\pm4.5$\\
10.65&$59.8^{+10.7}_{-5.6}$&$39.1^{+8.0}_{-3.0}$&$20.6^{+6.6}_{-4.9}$&$43.2^{+11.0}_{-6.4}$&$26.1^{+8.5}_{-4.6}$&$17.1^{+6.0}_{-3.7}$&$202.4\pm5.9$&$92.4\pm4.0$&$110.0\pm4.3$\\
10.75&$53.3^{+6.2}_{-3.3}$&$43.1^{+5.9}_{-3.4}$&$10.2^{+3.5}_{-2.9}$&$33.5^{+6.1}_{-2.3}$&$28.5^{+5.4}_{-2.6}$&\,\,\,$5.0^{+3.3}_{-1.5}$&$148.5\pm5.0$&$77.2\pm3.6$&$71.2\pm3.5$\\
10.85&$56.1^{+5.8}_{-6.6}$&$44.6^{+4.9}_{-4.2}$&$11.5^{+6.4}_{-4.2}$&$41.5^{+5.7}_{-4.1}$&$34.6^{+4.2}_{-3.5}$&\,\,\,$6.9^{+5.2}_{-3.0}$&$128.3\pm4.7$&$69.5\pm3.4$&$58.8\pm3.2$\\
10.95&$34.5^{+6.9}_{-6.6}$&$31.8^{+6.6}_{-6.2}$&\,\,\,$2.7^{+1.0}_{-1.1}$&$18.3^{+5.6}_{-4.3}$&$16.6^{+5.0}_{-4.0}$&\,\,\,$1.7^{+1.5}_{-1.1}$&$85.3\pm3.8$&$53.4\pm3.0$&$31.9\pm2.3$\\
11.05&$33.4^{+2.5}_{-2.7}$&$31.1^{+2.0}_{-1.8}$&\,\,\,$2.3^{+1.2}_{-1.6}$&$19.9^{+3.2}_{-3.2}$&$19.3^{+2.8}_{-3.3}$&\,\,\,$0.6^{+0.7}_{-0.6}$&$57.2\pm3.1$&$39.2\pm2.6$&$18.0\pm1.8$\\
11.15&$12.7^{+6.5}_{-3.3}$&$11.6^{+4.8}_{-2.9}$&\,\,\,$1.1^{+1.2}_{-0.9}$&$10.7^{+5.6}_{-4.4}$&$10.0^{+5.0}_{-3.8}$&\,\,\,$0.6^{+0.4}_{-0.6}$&$40.6\pm2.6$&$29.4\pm2.2$&$11.2\pm1.4$\\
11.25&$11.8^{+2.5}_{-3.9}$&$11.4^{+2.2}_{-3.8}$&\,\,\,$0.5^{+0.4}_{-0.5}$&\,\,\,$9.1^{+2.3}_{-2.4}$&\,\,\,$8.7^{+2.1}_{-2.4}$&\,\,\,$0.5^{+0.3}_{-0.5}$&$16.8\pm1.7$&$13.1\pm1.5$&$3.7\pm0.8$\\
11.35&\,\,\,$7.3^{+2.8}_{-3.6}$&\,\,\,$7.3^{+2.8}_{-3.6}$&\,\,\,\,\,\,\,\,\,-\,\,\,\,\,\,\,\,\,&\,\,\,$5.5^{+2.6}_{-3.0}$&\,\,\,$5.5^{+2.6}_{-3.0}$&\,\,\,\,\,\,\,\,\,-\,\,\,\,\,\,\,\,\,&$11.3\pm1.4$&$10.0\pm1.3$&$1.4\pm0.5$\\
11.45&\,\,\,$4.3^{+1.9}_{-1.6}$&\,\,\,$3.7^{+2.1}_{-1.9}$&\,\,\,$0.6^{+0.5}_{-0.6}$&\,\,\,$4.3^{+2.0}_{-1.6}$&\,\,\,$3.7^{+2.1}_{-1.9}$&\,\,\,$0.6^{+0.5}_{-0.6}$&$3.9\pm0.8$&$3.0\pm0.7$&$0.8\pm0.4$\\
11.55&\,\,\,$0.9^{+1.0}_{-0.9}$&\,\,\,$0.9^{+1.0}_{-0.9}$&\,\,\,\,\,\,\,\,\,-\,\,\,\,\,\,\,\,\,&\,\,\,$0.9^{+1.0}_{-0.9}$&\,\,\,$0.9^{+1.0}_{-0.9}$&\,\,\,\,\,\,\,\,\,-\,\,\,\,\,\,\,\,\,&$2.0\pm0.6$&$2.0\pm0.6$&\,\,\,\,\,\,\,\,\,-\,\,\,\,\,\,\,\,\,\\
11.65&\,\,\,\,\,\,\,\,\,-\,\,\,\,\,\,\,\,\,&\,\,\,\,\,\,\,\,\,-\,\,\,\,\,\,\,\,\,&\,\,\,\,\,\,\,\,\,-\,\,\,\,\,\,\,\,\,&\,\,\,\,\,\,\,\,\,-\,\,\,\,\,\,\,\,\,&\,\,\,\,\,\,\,\,\,-\,\,\,\,\,\,\,\,\,&\,\,\,\,\,\,\,\,\,-\,\,\,\,\,\,\,\,\,&$0.7\pm0.3$&$0.5\pm0.3$&$0.2\pm0.2$\\
11.75&\,\,\,\,\,\,\,\,\,-\,\,\,\,\,\,\,\,\,&\,\,\,\,\,\,\,\,\,-\,\,\,\,\,\,\,\,\,&\,\,\,\,\,\,\,\,\,-\,\,\,\,\,\,\,\,\,&\,\,\,\,\,\,\,\,\,-\,\,\,\,\,\,\,\,\,&\,\,\,\,\,\,\,\,\,-\,\,\,\,\,\,\,\,\,&\,\,\,\,\,\,\,\,\,-\,\,\,\,\,\,\,\,\,&$0.2\pm0.2$&$0.2\pm0.2$&\,\,\,\,\,\,\,\,\,-\,\,\,\,\,\,\,\,\,\\
\hline
\end{tabular}
\end{adjustwidth}
\end{table*}

\begin{figure}
\resizebox{\hsize}{!}{\includegraphics{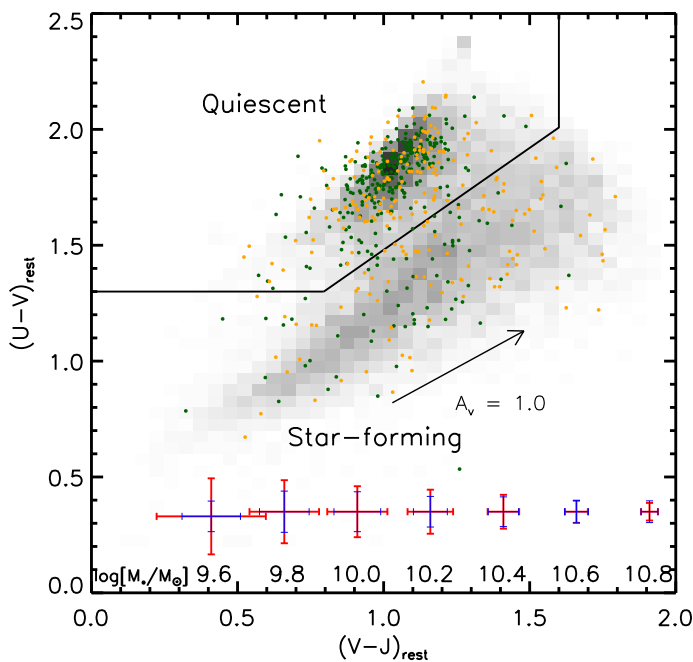}}
\caption{Rest-frame $U-V$ versus $V-J$ diagram for galaxies compiled from all clusters, with stellar masses $M_{\star}\geq 10^{10}\,\mathrm{M_{\odot}}$ and within projected $R \leq 1000\,\mathrm{kpc}$. \textit{Green:} Spectroscopic cluster members with $|\Delta z_{\mathrm{spec}}| \leq 0.02$. \textit{Orange:} Photometric cluster members with $|\Delta z_{\mathrm{phot}}| \leq 0.08$. \textit{Grey distribution:} UltraVISTA field galaxies with redshifts $1.0<z<1.4$ in the same mass range. The error bars present typical uncertainties at the depth of our photometry for different stellar masses, and separately for quiescent (red) and star-forming (blue) galaxies.}
\label{fig:UVJ}
\end{figure}

We measure rest-frame magnitudes in different bands based on the best-fit SEDs. In this study we use the rest-frame $U-V$ and $V-J$ colours to separate star-forming from quiescent galaxies, which is shown to work well even in the presence of dust reddening \citep[e.g.][]{wuyts07, williams09, patel12}. 
The SEDs are taken from a dedicated \texttt{EAZY} run, where, only for the purpose of measuring rest-frame colours, the redshifts of all galaxies are fixed to the cluster mean redshift. Figure \ref{fig:UVJ} shows the rest-frame colour distribution of galaxies with stellar masses exceeding $10^{10}\,\mathrm{M_{\odot}}$ and projected distances $R <  1000\,\mathrm{kpc}$ from any of the cluster centres. 

We note that there are small offsets between the quiescent loci in the rest-frame $UVJ$ colour distribution between the different clusters, and when compared to the COSMOS/UltraVISTA reference field. Similar trends were found in several previous studies \citep{whitaker11,muzzin13b,skelton14,leebrown17,vdB18}, and this suggests some residual uncertainties in the photometric calibration. In this study, we manually shift the $UVJ$ colour distributions back to the distribution from the COSMOS/UltraVISTA field in the redshift range $1.0<z<1.4$, by applying offsets that re-align the quiescent loci between different studies. The mean absolute shifts applied are 0.05 in both $U-V$ and $V-J$. 

After inspecting the bimodal galaxy distribution by eye, we select a sample of quiescent galaxies following the criteria:
\begin{equation}
 \mathrm{U-V} > 1.3 \,\,\,\cap\,\,\, V-J < 1.6 \,\,\,\cap\,\,\, U-V > 0.60+(V-J), 
\end{equation}  
which are close to the criteria used in \citet{muzzin13b} for the UltraVISTA sample.

Our analysis relies on the ability to separate star-forming from quiescent galaxies based on their $U-V$ and $V-J$ rest-frame colours\footnote{How galaxies evolve in their $UVJ$ colours depends on their star formation histories, and particularly on the way in which they quench \citep[e.g.][]{belli19}. As noted in e.g.~\citet{leja19b}, this choice of rest-frame colours does not necessarily well separate galaxies with low amounts of residual star formation from those that are truly ``dead''. Even though we perform exactly the same selection on the field and cluster galaxies, our results need to be regarded with this caveat in mind.}. To estimate the effect photometric uncertainties have on this selection, we take 50 Monte Carlo realisations based on our photometric catalogues, where we perturb the aperture fluxes within their estimated uncertainties following a normal distribution. We estimate rest-frame colours for the galaxies in each perturbed catalogue, and study the standard deviation of the results. The error bars in the lower part of Fig.~\ref{fig:UVJ}  show the median uncertainties in $U-V$ and $V-J$ separately, at different stellar masses. Based on this experiment, we estimate a net effect on the numbers of quiescent and star-forming cluster galaxies of less than 10\% ($<$0.05 \texttt{dex}), even at the lowest stellar masses considered in this work. Since the intrinsic colour distributions were already smeared and broadened by measurement uncertainties before we added extra noise, the true effect is likely smaller than this estimate. Since the inferred bias is small compared to other sources of uncertainty we consider, we do not attempt to correct for this effect.

\subsection{Completeness correction \& Richness measurements}\label{sec:completeness}
To characterise the completeness of the sources detected from the $K_\mathrm{s}$-band stacks, we measure the recovery rate of mock sources that were added to the science images. For this experiment, identical detection parameters were used as for the construction of the photometric catalogues. All sources we inject have an exponential (i.e.~S\'ersic=1) light profile with half-light radii in the range 1-3~kpc (uniformly distributed), ellipticities in the range 0.0 to 0.2 (uniformly distributed), and cover a wide range of magnitudes (uniformly distributed between 15 and 28) around the detection threshold. We injected $\sim30\,000$ galaxies per cluster, spread over 60 runs to not significantly affect the overall properties of the images with those simulated sources. To perform a proper completeness correction, the correction factors are dependent on the intrinsic magnitude distribution \citep[to account for Eddington bias][]{eddington1913,teerikorpi04}. We do take this correction into account, but, as illustrated in Appendix~\ref{app:eddington}, this has a minimal impact on our results. The PSF of the $K_\mathrm{s}$-band stacks (Image Quality reported in Table~\ref{tab:dataoverview}) are taken into account when adding the sources. 

The limiting magnitudes that are reported in Table~\ref{tab:dataoverview} correspond to the magnitude limit at which 80\% of the mock sources are still detected. Stellar mass limits corresponding to these magnitude limits are also reported in the Table. These are based on a single burst stellar population \citep[template from][]{bc03} formed at $z_{\mathrm{form}}=3.0$, with a \citet{chabrier03} IMF, no dust and solar metallicity. We note that younger stellar populations (such as those in star-forming galaxies) are brighter at the same stellar mass, and we assume that their stellar mass limit is 0.2 \texttt{dex} below that for quiescent galaxies \citep{bell01}.

To be able to scale the galaxy counts fairly in the SMF stack, even at low stellar masses when not every cluster is complete, we define a richness parameter $\lambda$. In this work, $\lambda$ is defined as the number of cluster galaxies, irrespective of galaxy type, with stellar mass $M_{\star} \geq 10^{10.2}\,\mathrm{M_{\odot}}$ measured within an $R<1000\,\mathrm{kpc}$ aperture. To account for foreground and background interlopers in these richness estimates, we perform a statistical subtraction of field galaxies from the COSMOS/UltraVISTA survey (accounting for different filter sets and depths compared to the cluster fields, a method that is also described in Appendix~\ref{app:approach2}). The resulting richnesses are listed in Table~\ref{tab:dataoverview}.

\subsection{Membership selection}\label{sec:membership}
When measuring the SMF of cluster galaxies (results presented in Sect.~\ref{sec:SMF}), it is important to account for line-of-sight interlopers. Ideally the identification of cluster members is fully based on spectroscopically measured redshifts of all sources found in the direction of a galaxy cluster. However, since the clusters are situated at high redshift, and given the low-mass galaxies we wish to study, this is practically impossible within a reasonable amount of telescope time. We will thus determine membership of sources that were \textit{not} targeted spectroscopically, based on our multi-band photometry, in combination with spectroscopic information of similar sources that \textit{were} targeted.

It is essential to define what ``similar'' means in this context. We have to separate the galaxy population between star-forming and quiescent galaxies, and further consider galaxies as a function of stellar mass and projected separation from the cluster centres. These three dimensions are expected to be important in the selection of spectroscopic targets, the photometric-redshift performance, and the success rate of measuring reliable spectroscopic redshifts.

\begin{figure}
\resizebox{\hsize}{!}{\includegraphics{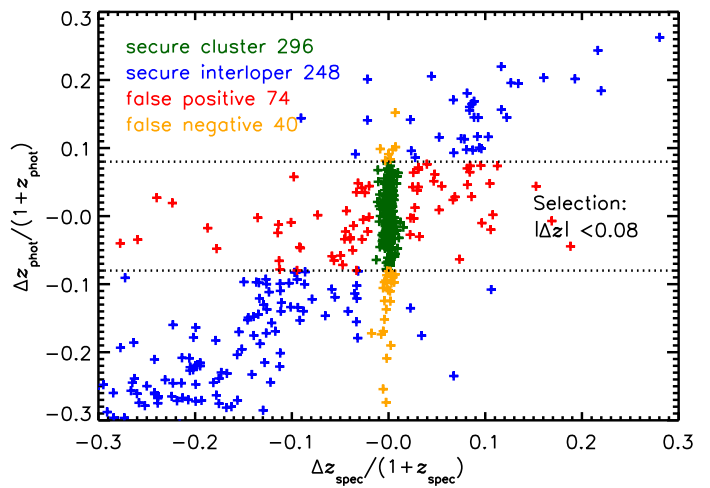}}
\caption{The same information as in Fig.~\ref{fig:speczphotz}, but here the x-axis shows the difference in $z_{\mathrm{spec}}$ with respect to the cluster redshift. This way we can identify spectroscopic cluster members (orange and green), as well as photometric cluster galaxies (red and green). Our fiducial measurement approach subtracts fore- and background interloper galaxies based on these relative numbers, after splitting the sample by galaxy type, and selecting similar sources in terms of stellar mass and projected radial distance from the cluster centers (cf.~\ref{fig:spectargets}).}
\label{fig:vdB13_plot_all}
\end{figure}

Our approach, which is comparable to that followed in \citet{vdB13}, relies on the spectroscopic subset being representative of the photometrically selected galaxy population. While this was a fundamental design goal of the GOGREEN targeting strategy, we test this assumption in Appendix~\ref{app:spectargets}, and find that it is valid. The approach is visualised in Fig.~\ref{fig:vdB13_plot_all}, which shows the same information as in Fig.~\ref{fig:speczphotz}, but here both axes are referenced with respect to the cluster mean redshift. As depicted in Fig.~\ref{fig:vdB13_plot_all}, the 11 clusters are essentially folded on top of each other. Spectroscopic cluster members (here these are defined as those for which $|\Delta z_{\mathrm{spec}}| \leq 0.02$, so that we are still probing cluster members that are $2-3\sigma_{\mathrm{los}}$ away from the mean redshift of the most massive GOGREEN clusters, cf.~Biviano et al., in prep.), and photometric cluster members (those for which, in the current example, $|\Delta z_{\mathrm{phot}}| \leq 0.08$) are marked  with different colours. 

In practice, for each un-targeted source, we estimate its membership probability based on the five most similar galaxies (in terms of radial distance and $M_{\star}$) that \textit{were} targeted. Targeted galaxies are divided in four classes, which follow the colours used in Fig.~\ref{fig:vdB13_plot_all}: ``secure cluster'',  ``secure interloper'', ``false positives'', and ``false negatives''. The membership correction factor $\mathrm{Corr}_i$ for each un-targeted galaxy $i$ is
\begin{equation}
\mathrm{Corr}_i  = \dfrac{\mathcal{N}\mathrm{(secure\,\,  cluster)} + \mathcal{N}\mathrm{(false\,\, negative)}    }{ \mathcal{N}\mathrm{(secure\,\,  cluster)} + \mathcal{N}\mathrm{(false\,\, positive)}     }, 
\end{equation}
where the $\mathcal{N}\mathrm{(X)}$ terms are the numbers of ``secure cluster'',  ``secure interloper'', and ``false positives'' among those five most-similar targets (cf.~lower panels of Fig.~\ref{fig:spectargets}).

This membership correction factor does not just range from 0 to 1, but also accounts for sources that were not even selected by their photometric redshifts. These ``false negatives'' can increase the weight to a value exceeding 1. We measure and assign such a membership weight for each un-targeted galaxy, to provide a statistical census of all cluster members. 

Correction factors are around unity when the number of ``false negatives'' and ``false positives'' are similar, and they become larger or smaller based on the chosen $|\Delta z_{\mathrm{phot}}|$ cut. We tested that the final results are not sensitive to the choice of $|\Delta z_{\mathrm{phot}}|$ (within reasonable limits, as also visualised in Fig.~\ref{fig:ellipses}), strengthening our confidence in this approach. We also performed an analysis where correction factors were measured in bins of stellar mass (instead of picking five similar galaxies per un-targeted galaxy). The results are very similar.

As a robustness check, we measure the SMF of quiescent galaxies by following an alternative approach which does not rely on the representativity of the spectroscopic sample. Rather, it subtracts the line-of-sight interlopers statistically by making use of a reference field; the COSMOS/UltraVISTA DR1 survey \citep{muzzin13a}. Here, only a subset of 13 filters is used from the entire DR1 catalogue (those filters listed in Table~\ref{tab:photometry}), and the entire analysis is performed identically to the GOGREEN analysis itself. This robustness check is most valuable for the quiescent galaxy population, for which spectroscopic redshift measurements are difficult and sparse at the low-mass end. The result based on this method is presented, and compared to that based on our fiducial method, in Appendix~\ref{app:approach2}; the results are fully consistent, which provides credibility to both approaches. We note that a statistical field subtraction of blue galaxies is non-informative due to the very low over-density of blue cluster galaxies against the fore- and background.

\section{The stellar mass function}\label{sec:SMF}
We measure the SMF of the GOGREEN cluster galaxies by considering all galaxies projected within 1000 kpc from the cluster centres (BCG positions), and applying the correction factors described in Sect.~\ref{sec:membership}. Even though the cluster sample covers a range of cluster masses, we note that 1000 kpc corresponds to a typical value of $R_{200}$ (Biviano et al., in prep.). Given this, whether apertures are chosen in fixed physical units (as we have chosen), or whether they are scaled with $R_{200}$, would not affect our results.

The photometry in the cluster fields has variable depth, leading to stellar mass detection limits that vary by several 0.1 \texttt{dex} between clusters (cf.~Table~\ref{tab:dataoverview}). We assign to each galaxy $i$, with stellar mass $M_{\star,i}$ and magnitude $K_\mathrm{s},\mathit{i}$, a total weight $w_i$, described as:
\begin{equation}\label{eq:weights}
w_i(M_{\star,i})=  \dfrac{1}{\mathrm{Compl(\mathit{K}_{s,\mathit{i}}} )}        \times            \mathrm{Corr}_i      \times   \dfrac{\sum_{cl} \lambda_{cl}}{\sum_{cl, M_{\star,i} > M_{\star,lim, cl}} \lambda_{cl}},
\end{equation}
where the first term corrects for sources that are undetected because they are too faint in the $K_\mathrm{s}$-band (as described in Sect.~\ref{sec:completeness} and Appendix \ref{app:eddington}). The second term corrects for cluster membership (cf.~Sect.~\ref{sec:membership}). Since we do not consider sources below the 80\% stellar mass detection limit of each cluster, the third term corrects for clusters that are missed because they do not allow one to probe galaxies at stellar mass $M_{\star,i}$. The numerator is a sum of the richnesses of all clusters; the denominator is a sum of the richnesses of clusters that are still complete at the stellar mass $M_{\star,i}$.
Richnesses are measured within the $R<1000\,\mathrm{kpc}$ aperture, and for galaxies that are sufficiently massive to be securely detected in all fields (cf.~Sect.~\ref{sec:completeness}). 
Applying such weights to each galaxy, we measure the cluster galaxy SMF down to $10^{9.5}\,\mathrm{M_{\odot}}$ ($10^{9.7}\,\mathrm{M_{\odot}}$) for star-forming (quiescent) galaxies; seven out of the 11 clusters are complete all the way down to these limits. 

\begin{figure*}
\resizebox{\hsize}{!}{\includegraphics{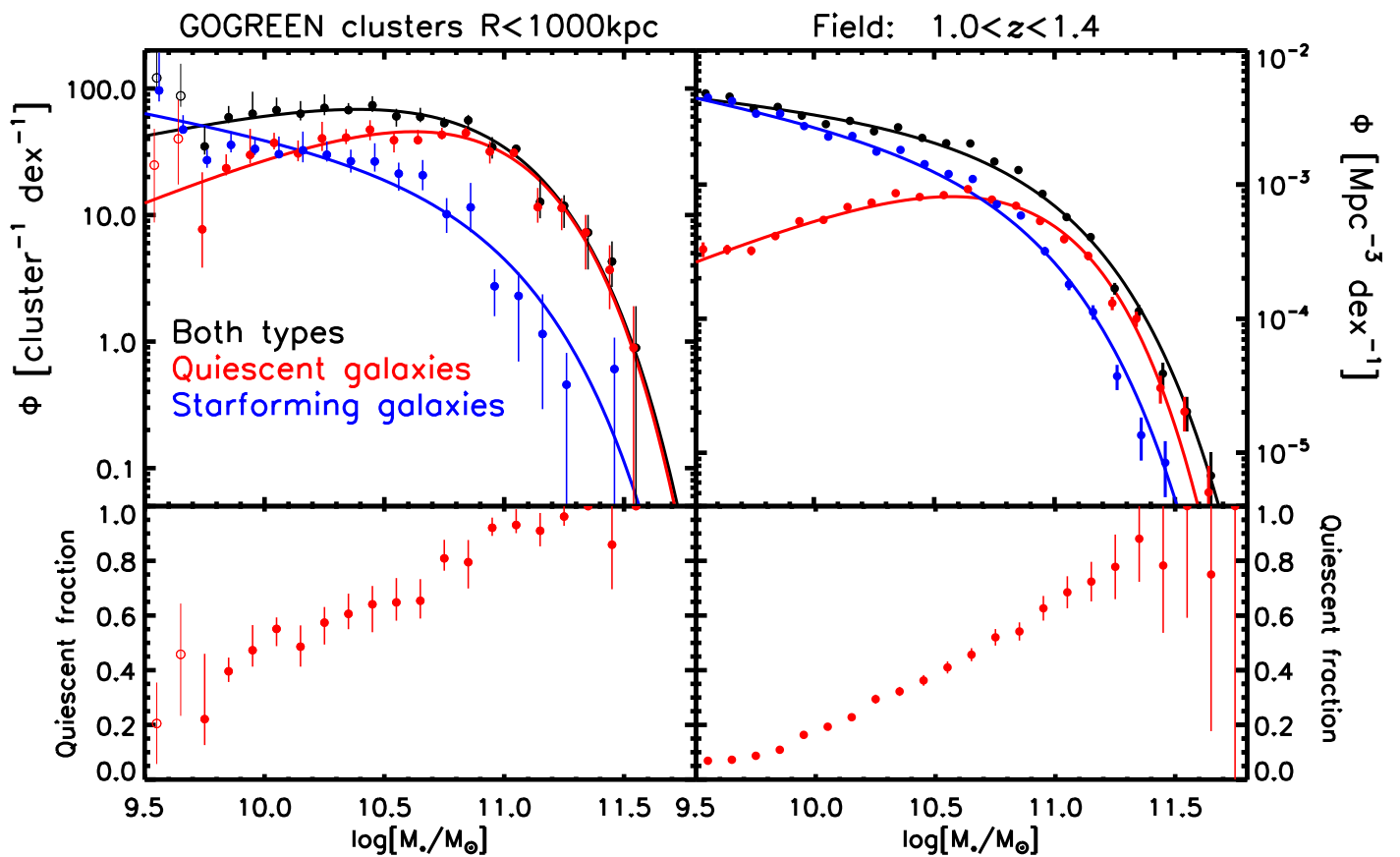}}
\caption{\textit{Top left panel:} SMF of cluster galaxies within $R \leq 1000\,\mathrm{kpc}$ from the cluster centres. \textit{Black points:} Total galaxy population. \textit{Blue and red data points:} The population of star-forming and quiescent galaxies, respectively. Small horizontal offsets have been applied compared to the black points for better visibility. The open circles mark points below the $80\%$ mass completeness limit, and, even though we perform an incompleteness correction, these are, conservatively, not used in the fitting. \textit{Top right panel:} SMF of co-eval field galaxies from the COSMOS/UltraVISTA survey ($1.0<z<1.4$). The best-fitting Schechter functions are included in both top panels. \textit{Lower panels:} Relative fraction of quiescent galaxies as a function of stellar mass, in the different environments.}
\label{fig:SMF_clustervsfield_raddist1000}
\end{figure*}

To probe the uncertainties on the SMF measurement, we consider cluster to cluster variations. We probe this source of uncertainty by performing the analysis on 100 bootstraps taken from the original cluster sample, where each time we draw 11 clusters with replacement. The error bars we use range from the 16th to the 84th percentile of the 100 bootstrap draws, and thus represent this source of uncertainty. In the hypothetical case where each cluster is identical, this quoted uncertainty equals, by construction, the Poisson uncertainties associated with the galaxy counts in the stack.

\subsection{Results and Schechter fits}
The measured SMF of the galaxies in the GOGREEN clusters is shown in the upper left corner of Fig.~\ref{fig:SMF_clustervsfield_raddist1000}, where galaxies with $R<1000$~kpc are considered. The data points are listed in Table~\ref{tab:datapoints}. At stellar masses $M_{\star} \gtrsim 10^{10}\,\mathrm{M_\odot}$, the abundance of quiescent galaxies exceeds that of star-forming galaxies (see also the lower panel, where the quenched fraction is plotted).

Following common practice, we model the SMF by fitting a Schechter \citep{schechter76} function to the data. This function is parameterized as 
\begin{equation}
\Phi(M)= \ln (10) \Phi^{*}\left[ \dfrac{M}{M^{*}}\right]^{(1+\alpha)} \exp\left[ -\dfrac{M}{M^{*}} \right],
\end{equation}
where $M^{*}$ is the characteristic mass, $\alpha$ the low-mass slope, and $\Phi^{*}$ the normalisation. We estimate the parameters $M^{*}$  and $\alpha$, which define the shape of the Schechter function, following the maximum likelihood approach described by Eq.~1~\&~2 in \citet{malumuthkriss86}. The un-binned data points are used for the fit, and following \citet{annunziatella14} \& \citet{vdB18}, we include weights for each galaxy to account for incompleteness and membership (cf.~Eq.~\ref{eq:weights}). The normalisation of the Schechter function, $\Phi^{*}$, is defined such that the integral over the considered stellar mass range (i.e.~stellar masses larger than $10^{9.5}\,\mathrm{M_\odot}$ or $10^{9.7}\,\mathrm{M_\odot}$) equals the number of all cluster galaxies (or more specifically, the sum of all weights).

The best fit parameters are listed in Table~\ref{tab:Schechter}, where two sources of uncertainty are quoted. The former are formal statistical uncertainties from the likelihood fit. The latter uncertainties indicate the range from the 16th to the 84th percentile of the best-fit parameters based on the 100 bootstrap samples, where each time 11 clusters were drawn with replacement. The best-fit Schechter functions provide good descriptions of the data (GoF, as defined and listed in Table~\ref{tab:Schechter}, are around unity), hence we do not consider a more complex fitting form such as a double Schechter function in this work. We note that, while there is a degeneracy between the best-fit Schechter parameters, we report uncertainties that are marginalised over the other two parameters.

\begin{figure*}
\resizebox{\hsize}{!}{\includegraphics{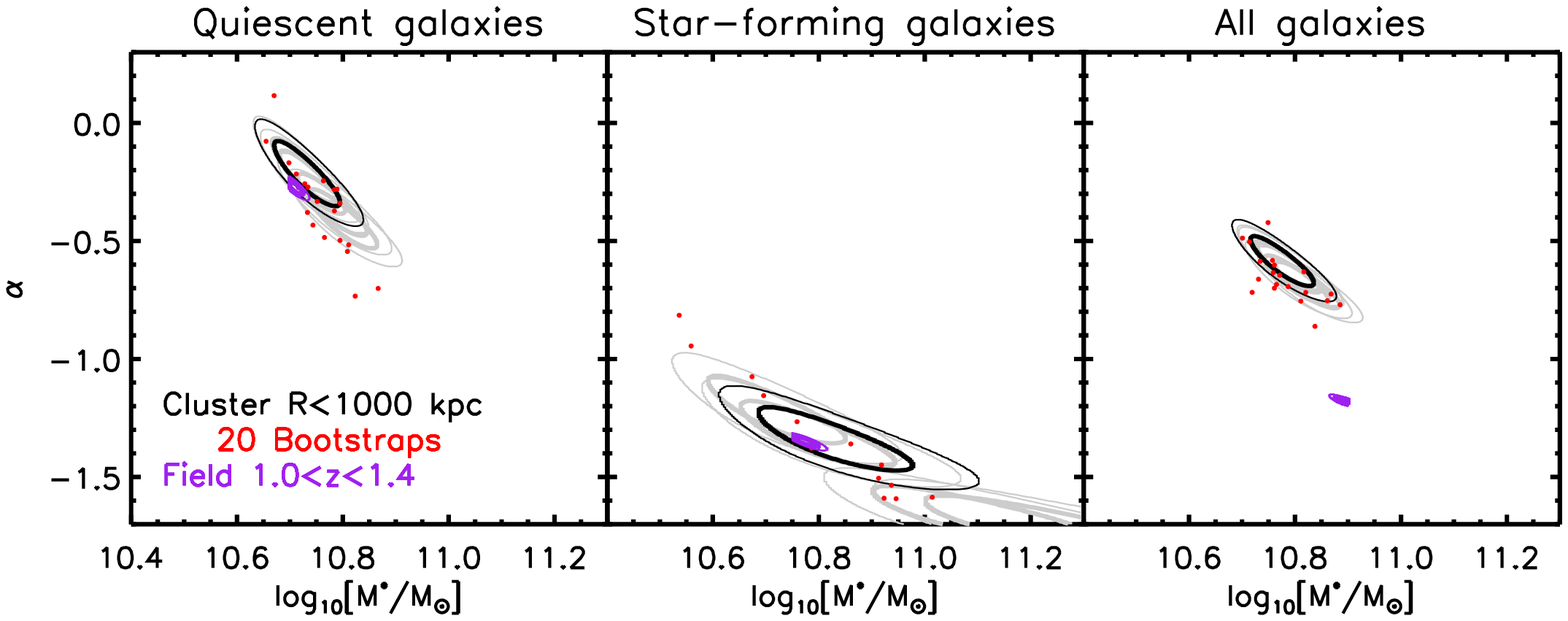}}
\caption{Comparison of the best-fitting Schechter parameters between cluster and field, when the galaxy population is divided between different galaxy types. \textit{Black contours:} 1- and 2-$\sigma$ uncertainties corresponding to the main analysis. \textit{Grey:} Robustness tests based on different initial photometric selection of cluster members. \textit{Red dots:} Results based on cluster bootstrap samples, where clusters are drawn with replacement. \textit{Purple contours:} 1- and 2-$\sigma$ uncertainties for the co-eval reference field.}
\label{fig:ellipses}
\end{figure*}

To illustrate this degeneracy, Fig.~\ref{fig:ellipses} shows the 68 and 95\% confidence regions around the two parameters that describe the shape of the Schechter function; $M^{*}$  and $\alpha$. In addition, 20 of the bootstrap values are shown (only the peaks of the respective likelihoods). The grey ellipses are uncertainty regions corresponding to the best-fit parameters obtained from an analysis with a different initial selection based on $z_{\mathrm{phot}}$. Whereas the solid black contours show the results for a fiducial selection of $|\Delta z_{\mathrm{phot}}| \leq 0.08$, the grey contours show results for $|\Delta z_{\mathrm{phot}}| \leq 0.04$, 0.06, 0.10, and 0.12. The ellipses all overlap with each other, which indicates that, as long as the interlopers are well characterised and accounted for, the results do not depend on the initial selection of galaxies (within reasonable limits). 

\subsection{Field comparison}
To be able to isolate the influence of the cluster environment on the galaxy population at these redshifts, we perform a comparison with the co-eval field galaxy population as probed by the COSMOS/UltraVISTA survey. We select all galaxies with photometric redshifts in the range $1.0<z<1.4$ in the unmasked area of the 1.62 deg$^2$ DR1 catalogue \citep{muzzin13a}, down to stellar masses of $10^{9.5}\,\mathrm{M_{\odot}}$. We note that significant over-densities have been identified at different redshifts in the COSMOS field \citep[e.g.][]{kovac10,laigle16,wang16,darvish17,darvish20}. Cosmic (or field-to-field) variance \citep[e.g.][]{somerville04} may thus also bias the probed galaxy population in the redshift interval $1.0<z<1.4$, between this field and the universe as a whole. We estimate the effect of cosmic variance on the measured field SMF based on the recipe described in \citep{moster11} for the boundaries of our survey, finding that it relative cosmic variance ranges from $\sim 5\%$ to $\sim 10\%$ for the lowest and highest mass galaxies we study in this volume. Since such a systematic uncertainty on the field comparison sample does not affect any of the conclusions drawn in this work, we do not explicitly take this variance into account in this analysis.

For this field study, in contrast to when we used the COSMOS/UltraVISTA for a statistical background correction in Sect.~\ref{sec:membership}, we use the full DR1 data set, which contains photometry in 30 filters. Since, at this depth, we are approaching the detection limit of the survey ($K_\mathrm{s}$-band magnitude completeness of 23.4 at 90\% detection rate), we have to make two corrections to the galaxy counts. Firstly, we note that luminous sources (or those with a high stellar mass) are detectable up to higher redshifts compared to those of lower mass. We therefore perform a ``$1/\mathrm{V}_{max}$ correction'' such as described in Sect.~3.4.1 of \citet{muzzin13b} (and references therein). Given the highest redshift, $z_{max}$, at which galaxies with stellar masses $M_{\star}$ can be securely ($>90\%$ completeness) detected, we define $\mathrm{V}_{max}$ to be the volume spanned by the COSMOS/UltraVISTA survey, from redshift 1.0 to $z_{max}$. Each source is then assigned a weight $\mathrm{V}_{tot}/\mathrm{V}_{max}$, where $\mathrm{V}_{tot}$ is total volume spanned by the COSMOS/UltraVISTA survey in the redshift interval $1.0<z<1.4$. All sources that have stellar masses lower than the 90\% completeness limit at that redshift are assigned a weight 0. Secondly, to account for residual incompleteness, we use the corrections estimated for the UltraVISTA detection band \citep[$K_\mathrm{s}$, Fig.~4 in][]{muzzin13a}. The products of these weights are included in the data points. They also enter in our maximum likelihood estimation, and we follow a similar procedure as for the cluster galaxy population.

The results are shown in the right panel of Fig.~\ref{fig:SMF_clustervsfield_raddist1000}, and the data points are listed in Table~\ref{tab:datapoints}. The best-fitting Schechter parameters are reported in Table~\ref{tab:Schechter} and visualised in Fig.~\ref{fig:ellipses}. We note that the results are similar to the best-fit Schechter parameters estimated by \citet{muzzin13b} based on the same data set, but in the redshift range $1.0<z<1.5$.

\begin{figure*}
\resizebox{\hsize}{!}{\includegraphics{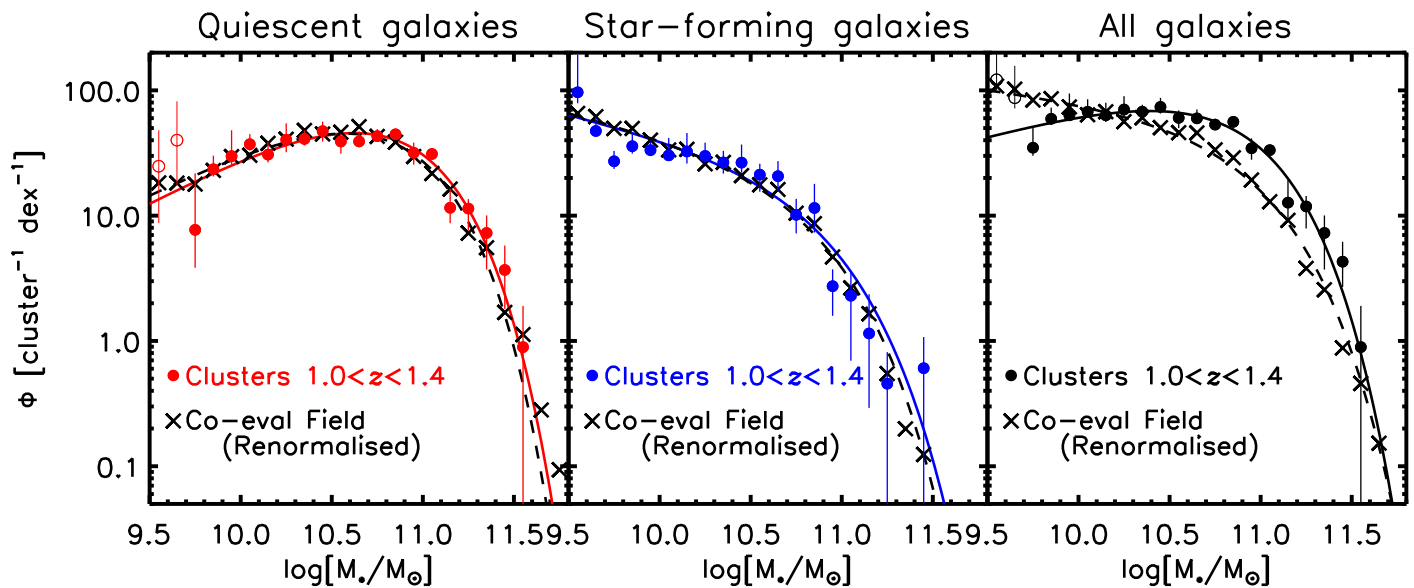}}
\caption{\textit{Left:} The SMF function of quiescent galaxies in the cluster environment ($R<1000$~kpc), compared to the field. The cluster data points are identical to those shown in Fig.~\ref{fig:SMF_clustervsfield_raddist1000}, while the field is normalised so that it integrates to the same number of quiescent galaxies down to $M_{\star} \geq 10^{9.5}\,\mathrm{M_{\odot}}$. \textit{Middle:} Same but for star-forming galaxies. The best-fitting Schechter functions are over-plotted. The resemblance in the shapes of the separate (quiescent versus star-forming galaxies) SMFs is evident. \textit{Right:} Same comparison but for all galaxies, where there clearly is a different SMF between cluster and field.}
\label{fig:SMF_comparetypes}
\end{figure*}

Several results are immediately apparent from Figs.~\ref{fig:SMF_clustervsfield_raddist1000}~\&~\ref{fig:ellipses}. Already in the redshift range $1.0<z<1.4$, galaxies in clusters have a significantly higher probability to be quenched than similarly massive galaxies in the field. 
Yet, if we consider quiescent galaxies only, the shape of the SMF of this galaxy type appears similar between cluster and field. The same is also true if we only consider star-forming galaxies. These points are illustrated more clearly in Fig.~\ref{fig:SMF_comparetypes}, where the cluster- and field SMF are shown in the same panels, this time divided by galaxy type. The field counts are normalised so that they integrate to the same number of galaxies down to $M_{\star} \geq 10^{9.5}\,\mathrm{M_{\odot}}$. Within the relatively small statistical uncertainties, the shapes of the distributions are essentially indistinguishable between cluster and field, for quiescent and star-forming galaxies (note the ellipses in Fig.~\ref{fig:ellipses}). 
On the contrary, the overall shape of the SMF of \textit{all} galaxies in the cluster and field environments is significantly different; there are relatively more low-mass galaxies in the field environment compared to the cluster environment (or, conversely, there are more massive galaxies in the cluster environment). These results are discussed in more detail in Sect.~\ref{sec:discussion}, but first, we quantify the contribution of the environment in the quenching of galaxies with another metric. 

\begin{table*}
\caption{Best-fitting Schechter parameters and their 1-$\sigma$/68\% confidence limits for different galaxy types in the cluster and field environments. In addition to the formal statistical uncertainty (first error) for the cluster data, we quote the bootstrap uncertainty (second error).}
\label{tab:Schechter}
\begin{center}
{\renewcommand{\arraystretch}{1.5}
\begin{tabular}{l l l l l l}
\hline
\hline
Environment & Type & $\mathrm{log_{10}[}M^{*}/\mathrm{M_{\odot}}]$ & $\alpha$ & $\Phi^{*a}$ &GoF$^b$\\
\hline
&All galaxies&$10.77^{+0.04+0.05}_{-0.04-0.04}$&$-0.59^{+0.07+0.08}_{-0.07-0.13}$&$64.65\pm 2.04^{+8.35}_{-6.82}$&0.91\\
$R<1000$~kpc&Quiescent galaxies&$10.73^{+0.04+0.06}_{-0.04-0.03}$&$-0.22^{+0.09+0.10}_{-0.09-0.21}$&$52.45\pm 2.16^{+4.65}_{-6.14}$&0.91\\
&Star-forming galaxies&$10.82^{+0.10+0.19}_{-0.09-0.12}$&$-1.34^{+0.09+0.19}_{-0.09-0.32}$&$10.31\pm 0.47^{+6.51}_{-5.38}$&1.38\\
\hline
&All galaxies&$10.80^{+0.05+0.06}_{-0.05-0.05}$&$-0.50^{+0.09+0.11}_{-0.09-0.18}$&$38.69\pm 1.66^{+5.36}_{-3.70}$&0.88\\
$R<500$~kpc&Quiescent galaxies&$10.78^{+0.05+0.08}_{-0.05-0.03}$&$-0.26^{+0.11+0.14}_{-0.10-0.30}$&$31.95\pm 1.67^{+4.45}_{-5.81}$&0.72\\
&Star-forming galaxies&$11.06^{+0.19+0.38}_{-0.16-0.33}$&$-1.53^{+0.11+0.30}_{-0.10-0.47}$&$2.41\pm 0.16^{+3.90}_{-1.86}$&1.48\\
\hline
&All galaxies&$10.89^{+0.01}_{-0.01}$&$-1.18^{+0.01}_{-0.01}$&$112.87\pm 0.78$& 2.60\\
Field&Quiescent galaxies&$10.70^{+0.01}_{-0.01}$&$-0.26^{+0.02}_{-0.03}$&$92.32\pm 1.24$& 1.36\\
&Star-forming galaxies&$10.77^{+0.02}_{-0.01}$&$-1.35^{+0.01}_{-0.02}$&$73.50\pm 0.59$& 2.34\\
\hline
\end{tabular} }
\end{center}
\begin{list}{}{}
\item[$^{\mathrm{a}}$]  Normalisation is reported as average per cluster, so in units [cluster$^{-1}$] for the cluster data, and [$10^{-5}$ Mpc$^{-3}$] for the reference field.
\item[$^{\mathrm{b}}$]  Even though maximum likelihood fits were performed on the unbinned data, we report goodness of fits (GoF) as $\chi^2/\mathrm{d.o.f.}$, where the best-fit models are compared to the binned data. For this we have assumed two-piece normal distributions for each data point, corresponding to the asymmetric uncertainties in Table~\ref{tab:datapoints}, where the $\pm 1\sigma$ range covers a 68\% total probability.
\end{list}
\end{table*}

\subsection{Quenched Fraction Excess}
A related measurement to the quenched fractions in clusters is that of the Quenched Fraction Excess ($QFE$), which describes the fraction of galaxies that would have been star-forming in the field, but are quenched by their cluster environment. Specifically, 

\begin{equation}\label{eq:QFE}
QFE= \frac{f_\mathrm{q,cluster}-f_\mathrm{q,field}} {1-f_\mathrm{q,field}},
\end{equation}
where $f_\mathrm{q,cluster}$ and $f_\mathrm{q,field}$ are the quenched fractions of galaxies in the cluster and field environment, respectively. The quenched fractions are a function of both stellar mass and environment, but whether $QFE$ is also a function of these parameters is a matter of debate, and this may depend on epoch/redshift and on exactly which environment is considered. 

We note that other terms are adopted to refer to a similar quantity as $QFE$, such as ``transition fraction'' \citep{vandenbosch2008}, ``conversion fraction'' \citep{balogh16,fossati17}, or ``environmental quenching efficiency'' \citep[e.g.][]{peng10,wetzel15,nantais17,vdB18}. We have adopted the terminology $QFE$, used in \citet{wetzel12} and \citet{bahe17}, since it seems intuitively closest to what is measured.

\begin{figure}
\resizebox{\hsize}{!}{\includegraphics{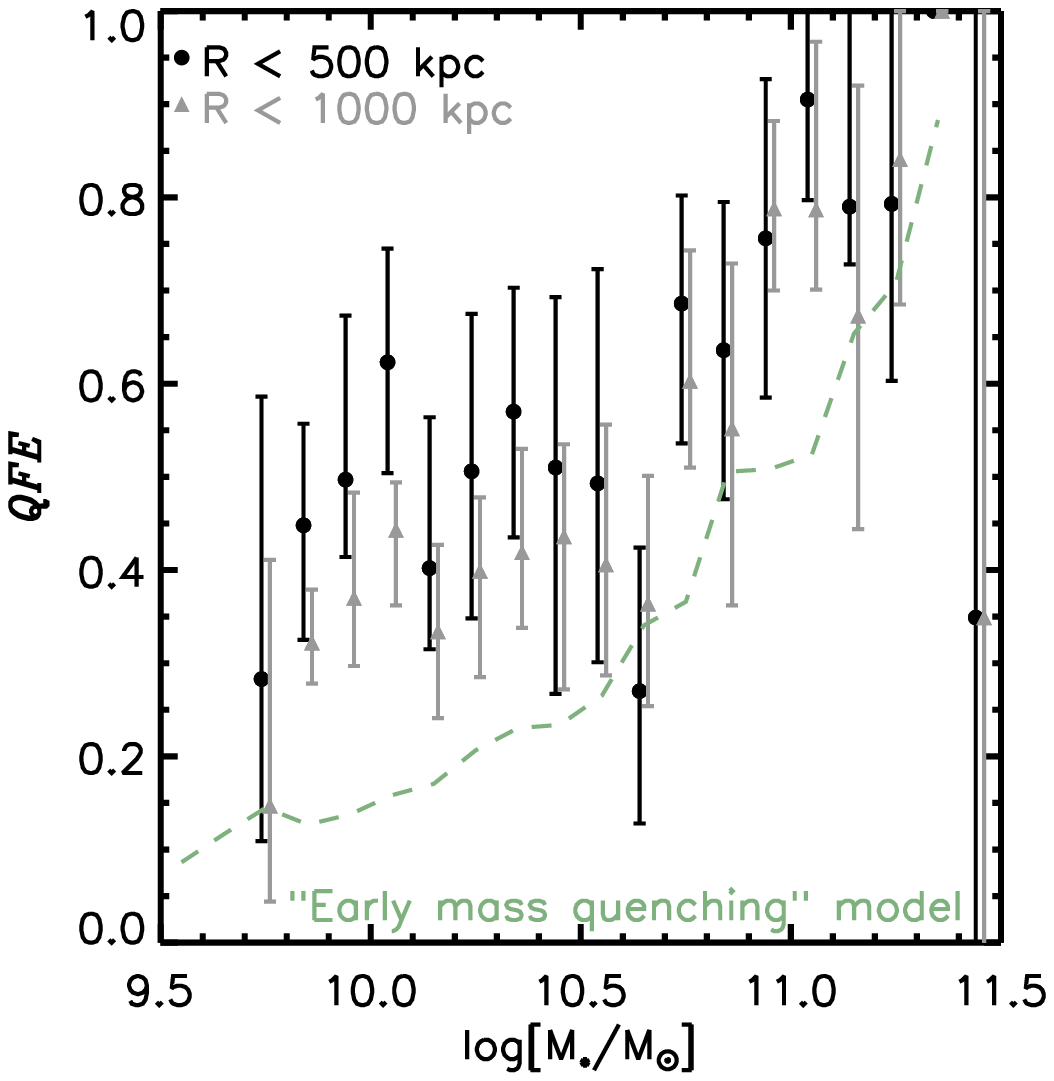}}
\caption{Quenched Fraction Excess ($QFE$) for cluster galaxies as a function of stellar mass. \textit{Black:} Considering cluster galaxies at $R<500$~kpc. \textit{Grey:} Considering cluster galaxies at $R<1000$~kpc. \textit{Green dashed:} Galaxy evolution model as described in Sect.~\ref{sec:discussion3}, where the field and cluster galaxies have started forming at different redshifts.}
\label{fig:EQE}
\end{figure}

In Fig.~\ref{fig:EQE} we present the $QFE$ of cluster galaxies as a function of stellar mass, and for the two different radial regimes: $R<500$~kpc and $R<1000$~kpc. Following the SMF measurements, we report errors that are estimated from the bootstrap resamplings. The effect of the environment is significant at all stellar masses ($QFE$ is well above zero over the entire range), and is even higher closer to the cluster centres ($R<500$~kpc) than when we also consider galaxies at larger projected radii \citep[cf.][]{vdB18,strazzullo19}. Furthermore, the $QFE$ is clearly dependent on stellar mass, with higher-mass galaxies having a higher probability to be quenched due to their environment.

\section{Discussion}\label{sec:discussion}
In this Section we first discuss our main results, which were presented in Sect.~\ref{sec:SMF}, at face value. Then, in subsections \ref{sec:discussion2}-\ref{sec:discussion1}, we discuss the status and predictions of a purely phenomenological model, as well as a more physically motivated model, of galaxy quenching. In these subsections, we discuss to what extent the measurements are reproduced by the models, and discuss where further tests and revisions may be required.

With the highly elevated quenched fractions measured for galaxies in the GOGREEN clusters, it is clear that these galaxies must have followed a different evolutionary path compared to those in the co-eval field. The substantial influence of a cluster environment in the quenching of galaxies does not come as a surprise, and is shown in many previous studies since \citet{dressler80} \citep[cf.~Fig.~7 in][for a compilation of a number of results in the literature]{nantais16}. What is more remarkable is the lack of an imprint this enhanced quenching process has on the separate SMFs of star-forming and quiescent galaxies.

Indeed, in the high-density environments probed in this work, there is no measurable difference in the shape of the SMF of \textit{star-forming} galaxies compared to the average field (cf.~Figs.~\ref{fig:ellipses}~\&~\ref{fig:SMF_comparetypes}). A similar result was found at lower redshift \citep[e.g.][]{peng10, vulcani13, vdB13, annunziatella14, annunziatella16} or at more-moderate over-densities at similar redshifts \citep{papovich18}. 
Thanks to our low detection limit of $10^{9.5}\,\mathrm{M_{\odot}}$ and high statistical precision due to the combined sample of 11 clusters, we can place very strong constraints on the similarity in the shape of the SMF of star-forming galaxies in different environments, compared to most previous studies. This high statistical precision is reflected by the relatively small uncertainties shown in Fig.~\ref{fig:ellipses}. For example, we find that there is a $\sim 10\%$ probability that the $\alpha$ parameter that describes the low-mass end of the star-forming SMF deviates by more than $\pm0.5$ from the best-fit field value. Furthermore, there is only a $\sim 10\%$ probability that the characteristic mass $M^{*}$ deviates by more than $0.30$ \texttt{dex} from the best-fit field value. These numbers are based on the bootstrapped cluster samples, and thus include cluster-to-cluster variance.

Remarkably, we also do not find a measurable difference between the SMF of \textit{quiescent} galaxies between GOGREEN clusters and the co-eval field studied in this work (cf.~Fig.~\ref{fig:SMF_comparetypes}). Here, again we showcase the precision of our measurement, by making a more stringent and quantitative statement regarding our finding that the SMF of quiescent galaxies has a similar shape between the cluster and field; we find that there is a mere $\sim 10\%$ probability that the $\alpha$ parameter that describes the low-mass end of the quiescent SMF deviates by more than $\pm0.3$ from the best-fit field value. Moreover, there is only a $\sim 10\%$ probability that the characteristic mass $M^{*}$ deviates by more than $0.12$ \texttt{dex} from the best-fit field value. Again, these numbers are based on the bootstrapped cluster samples, and thus include cluster-to-cluster variance. We note that \citet{chan19} obtain a similar result based on measurement of the rest-frame $H$-band luminosity function of red-sequence galaxies in seven of the GOGREEN clusters. 

At first glance, the similarity in the shape of the SMF of quiescent galaxies in different environments is surprising given the much-higher total quiescent fraction of galaxies in clusters compared to the field. In the local Universe, studies that measure an excess of low-mass quenched galaxies in high-density environments compared to in lower-density environments, attribute this to a different quenching mechanism at play \citep{peng10,bolzonella10,moutard18}. We note that there is still some debate concerning the impact of environment on the shape of the different SMFs in the local Universe. For instance, some studies that do not find a difference may be hampered by a too high stellar mass completeness limit, so that a potential trend may not be detectable in the data \citep[e.g.][]{vulcani13,calvi13}. 

In contrast, the \textit{total} SMF of galaxies in the clusters and the field (with stellar masses $M_{\star} \geq 10^{9.7}\,\mathrm{M_{\odot}}$) is radically different. A two-sample KS test \citep[e.g.~Chapter 14 in][]{numericalrecipes92} indicates that the probability that both samples of galaxies are drawn from the same parent distribution is $P\sim 10^{-21}$. While SMFs of individual galaxy types are similar in the different environments, the total SMF is not because of the different fractions of quenched galaxies in cluster and field.

It is worthwhile to point out that, when comparing our works to e.g.~\citet{kawinwanichakij17} and \citet{papovich18}, who study the influence of environment on the star-forming properties of galaxies in the ZFOURGE and NMBS surveys, we use a different definition of field. In the present work, we take the ``field'' to be an average/representative part of the Universe. This therefore includes numerous moderate over-densities like galaxy groups, which may trigger and/or enhance quenching. In contrast, some other studies define their lowest-density quartile as the basis compared to which environmental quenching processes are quantified \citep{peng10,kawinwanichakij17,papovich18}. Besides this aspect, we study massive galaxy clusters, whereas their relatively small survey area only probes more moderate galaxy densities. With both differences taken together, we measure the influence of the environment over a different range in environmental densities, and it is thus remarkable that we obtain qualitatively similar results as those studies. 

The substantially elevated quenched fraction of galaxies in clusters, in combination with the similarity in the SMFs of quiescent and star-forming galaxies, provides insights into \textit{how} quenching operates in these environments.

\subsection{The need for environmental quenching}\label{sec:discussion2}
While there is a clear need for environmental quenching to explain the quenched excess in the GOGREEN clusters, it is questionable whether similar processes are at play as in the local Universe. Here we discuss our results in the context of the quenching model that was introduced and employed by \citet{peng10}. The key feature of this model, which is supported by observations in the $z<1$ Universe, is that mass and environment affect the quenched fraction of galaxies in a way that is separable \citep[although some recent work challenges this picture,][]{darvish16,pintoscastro19}. This led Peng et al. to introduce concepts of mass- and environmental quenching. An important aspect of this model is that neither of these quenching modes affects the shape of the SMF of star-forming galaxies (as a function of time or environment). 
Our observation that the shape of the SMF of star-forming galaxies between the GOGREEN cluster galaxies and the co-eval field is similar, is thus in line with the \citet{peng10} model. One of the quenching processes that keeps the shape of the SMF of star-forming galaxies intact and unchanging with time, is a process that operates completely independently of stellar mass.
This is what \citet{peng10} refers to as environmental quenching.
In its basic form, this requires the $QFE$ to be constant as a function of stellar mass. In contrast to the local Universe, where the environment is indeed observed to have this effect \citep{baldry06,peng10,delucia12,phillips15}, this is clearly not the case for the GOGREEN clusters at $z\gtrsim 1$ \citep[cf.~Fig.~\ref{fig:EQE}\footnote{While the overall trend shown in Fig.~\ref{fig:EQE} is increasing with stellar mass, we can, with the current uncertainties not rule out that the $QFE$ plateaus at masses $\mathrm{log[M_\star/M_{\odot}]} \lesssim 10.5$, and only strongly increases for higher masses.}, also see][]{balogh16,kawinwanichakij17}. If the enhanced quenched fractions of galaxies in clusters were due to an environmental quenching process that was independent of stellar mass, this would have resulted in an over-abundance of quenched low-mass galaxies, resulting from the high abundance of star-forming galaxies that undergo quenching \citep[][]{papovich18}. 

\begin{figure}
\resizebox{\hsize}{!}{\includegraphics{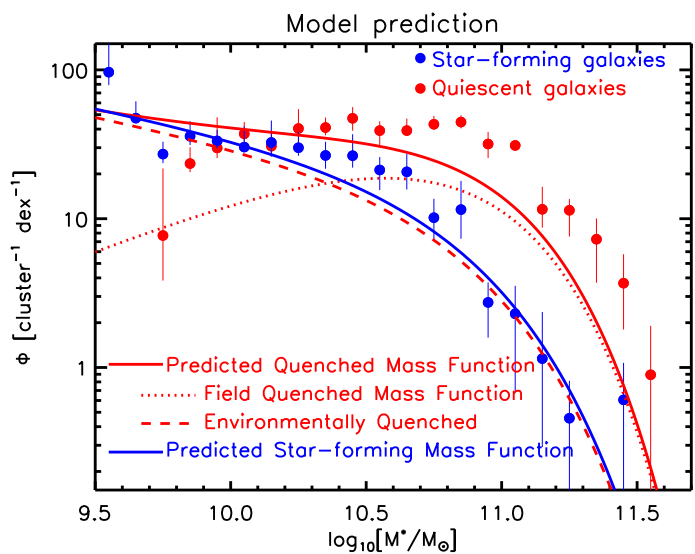}}
\caption{A comparison between the data and predictions from the model described in Sect.~\ref{sec:discussion2}. The main model assumption is that the quenched fraction excess is independent of stellar mass, and fixed to the best-fit value of $QFE=0.47\pm0.03$. This corresponds to the pure environmental quenching scenario from \citet{peng10}, but is contrary to the mass dependence we observe, cf.~Fig.~\ref{fig:EQE}. While the star-forming SMF is well reproduced by this model, there is a clear mis-match between the predicted (red solid line) and observed (red points) SMF of quiescent (cluster) galaxies.}
\label{fig:pengsimple}
\end{figure}

We stress this point more clearly in Fig.~\ref{fig:pengsimple}, where we explore the additional environmental quenching that is required to take place compared to the field\footnote{We remind the reader that the field, as defined in this work, is representative for the universe as a whole. It thus contains numerous smaller-scale over-densities, such as groups and filaments, where environmental quenching may also be occurring. Here, we quantify the excess quenching caused by the cluster environment, compared to this baseline.}, in order to match the quiescent distribution of galaxies observed in the clusters. For this, we take the Schechter functions fitted to the field galaxy populations of star-forming and quiescent galaxies as a starting point. 
Since the total number of galaxies is conserved in this model, the total normalisation of galaxies (red+blue) is set by the total number of galaxies contained in the data points (which represent the cluster population). The red dotted line represents the re-normalised fit to the field quiescent galaxy SMF, which represents the population of galaxies that have been intrinsically (=``mass'') quenched. On top of this, there is a certain fraction of blue field galaxies, described by the blue Schechter function, that are ``environmentally quenched'' and added to the quiescent population. This is represented by the dashed line, whose height is set by matching to the overall fraction of quenched galaxies. The best-fit model has a single value of $QFE=0.47\pm0.03$, and is shown by the solid red line. The single value of the QFE is a direct consequence of our assumption that environmental quenching in the \citet{peng10} picture is independent of stellar mass. In the redshift range we consider, this simple model fails at reproducing the data over the entire stellar-mass range. While this environmental quenching term explains the excess quenching of galaxies in local galaxy clusters, there must be an additional/different quenching mode that dominates at higher redshift. 

\subsection{The formation time of galaxies and the pace of galaxy evolution}\label{sec:discussion3}
The shape of the SMF of quiescent galaxies is indistinguishable between cluster and co-eval field (at least in the redshift range we study, $1.0<z<1.4$). It is therefore worthwhile to consider a single quenching process that may be responsible for quenching galaxies in both environments, and which acts like ``mass quenching'' in the \citet{peng10} framework.  Qualitatively, the quenched fractions of galaxies in the GOGREEN clusters are comparable to that measured for the field in the local ($z\lesssim 0.5$) Universe. Therefore, a simple explanation is one in which galaxies in clusters quench through the same processes as those in the field, but simply do so at an earlier time. We thus consider a scenario in which galaxies that are destined to become part of our clusters start their formation ``early'' with respect to galaxies in the field, but quench via a similar physical process.
In fact, in the experiment described in Sect.~6 of \citet{peng10}, it is assumed that there is a 1 Gyr delay in the formation of (seed) galaxies in the D1 compared to the D4 regions, where D1 and D4 are the lowest and highest density quartile, respectively. With a formation redshift of $z_{\mathrm{form}}=10$ for D4, this means a formation redshift of $z_{\mathrm{form}}\simeq 4$ for D1.
\begin{figure*}
\begin{center}
\resizebox{0.7\hsize}{!}{\includegraphics{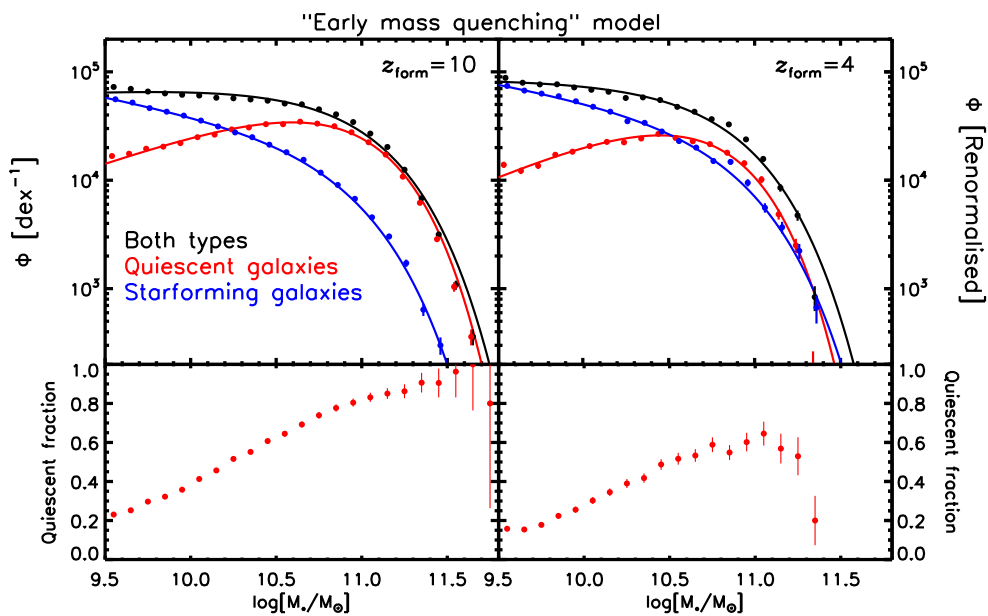}}
\caption{\textit{Top panels:} The result of a galaxy evolution models following the formalism for mass quenching, as described in \citet{peng10}. These assume a formation redshift of $z_{\mathrm{form}}=10$ (left panel) and $z_{\mathrm{form}}=4$ (right panel), and the panels compare the results at $z=1.2$. For easy comparison, the right panel was renormalised by a factor 5.6 so that both panels have the same number of galaxies in the plotted mass range. \textit{Lower panels:} Relative fraction of quiescent galaxies for the model runs with the two different formation redshifts. In this model we observe a qualitatively similar SMF of quiescent galaxies in both panels, as well as for the star-forming galaxies. Moreover, the mass-dependent $QFE$ of the $z_{\mathrm{form}}=10$ population compared to the $z_{\mathrm{form}}=4$ population is similar to what we observe (cf.~Fig.~\ref{fig:EQE}). Details on the model are given in Sect.~\ref{sec:discussion3} and in \citet{peng10}.}
\label{fig:SMF_earlymassquenchingmodel}
\end{center}
\end{figure*}

We attempt to redo the experiment described in \citet{peng10}, and make reasonable assumptions where information is missing. For instance, we start with a distribution of star-forming seed galaxies with masses between $10^2 - 10^9\,\mathrm{M_{\odot}}$\footnote{\citet{peng10} do not specify their mass range of seed galaxies. They mention the precise value of the high-mass cutoff is immaterial for the final results, as long as the cutof is at low-enough mass to avoid over-populating the initial population with very massive galaxies.} (with a mass distribution described by a power law with logarithmic slope $\alpha=-1.3$), and let them, in steps of 20 Myr, grow in stellar mass through in-situ star formation following the star forming main sequence as parametrized in \citet{schreiber15} \citep[we could have used the results from][the exact parametrization does not affect the result]{whitaker14,speagle14}. Galaxies are quenched with probabilities proportional to the instantaneous SFR which, given the relation between SFR and stellar mass, also results in a mass-dependent quenching probability. We confirm, as described in \citet{peng10}, that this builds Schechter-like distributions both for star-forming and quiescent galaxies. Interestingly, in this experiment, we find that the difference in formation time of 1 Gyr ($z_{\mathrm{form}}=10$ for the cluster galaxies versus $z_{\mathrm{form}}=4$ for the field galaxies) leads to a $QFE$ at $z=1.2$ that is qualitatively similar to what we observe in the GOGREEN clusters compared to the field (although this simplistic model, which we call ``early mass quenching'' suggests that we may need an even larger difference to reproduce the exact trend, cf.~Fig.~\ref{fig:EQE}). The resulting SMFs are shown in Fig.~\ref{fig:SMF_earlymassquenchingmodel}, and these indeed have forms that are qualitatively similar to the observations.

We note that such a difference in formation time between galaxies in cluster and field should express itself as a corresponding difference in the ages of the stellar populations of quiescent galaxies in clusters and in the field. Measurements like these, and their interpretation, is still a topic of debate \citep{vandokkum07,gobat08,saglia10,rettura10,cooper10b, lin16}. However, all those studies find, at fixed mass, age differences that are less than, or at most, 1 Gyr between cluster and field. Since the required difference in formation time is likely more than 1 Gyr to explain the measured $QFE$, this seems \textit{inconsistent with the measured ages} (also see Webb et al., in prep.).

Within the framework of this simple model, we consider another option, namely that the SFRs of galaxies are elevated in the environments that are progenitors to our clusters, compared to the co-eval field at those early times. Even though the environmental-dependence of the star-forming main sequence is still a topic of debate \citep[e.g.][]{vulcani10,popesso11,koyama13,paccagnella16,paulinoafonso18,wang18,tomczak19}, it is clear that there is, at the epoch of the observation, no large difference in the star-forming main sequence between cluster and field \citep[For the GOGREEN data set we measure an offset of 0.14 \texttt{dex} in the sSFR between cluster and field, at a significance of 3.1$\sigma$,][]{old20}. This has not necessarily been the case at earlier times. Indeed, some distant (proto-) clusters appear to be forming stars at particularly high rate \citep{casey15,wang16,oteo18,cheng19}. 
With an increased SFR of galaxies at early times, our naive expectation is that, at fixed formation epoch, this would increase the number of mass-quenched galaxies (as this quenching process is proportional to the SFR of individual galaxies). However, since the increased SFR would result in a proportional increase in the population of star-forming galaxies, we would notice no gain in the quenched \textit{fraction} in this experiment (the $QFE$ would be equal to zero at all masses). Therefore, \textit{the observed trend shown in Fig.~\ref{fig:EQE} is not reproduced.}

We note that the above experiment (``early mass quenching'') also includes mergers, and ``merger quenching'' as implemented by \citet{peng10}. In their model, when major mergers happen at $z<3$, they are assumed to quench the participating galaxies. We take the redshift-dependent major merger rates adopted by Peng et al.~for the D1 and D4 density quartiles\footnote{While we take the values adopted by \citet{peng10}, we note that the mass-, redshift- and environmental dependence of galaxy merger rates are still debated, and an active topic of research, cf.~\citet{rudnick12,delahaye17,duncan19}.}, and apply them independently of stellar mass (so we do not specifically estimate merger rates in cluster environments). In practise, in each step of 20 Myr, we randomly assign galaxies that are supposed to merge in this time interval. We then go through the mass-sorted list of galaxies that are flagged to merge, and pair them up so mergers happen between galaxies with most similar masses. We note that, within the limits and implementations we have tried, the inclusion of mergers have a minimal effect on the measured SMFs, nor do they result in a drastic quenched fraction excess in different environments. However, we note that \citet{tomczak17} perform a similar model to match to the galaxy SMF of the ORELSE galaxy clusters, but leave the merger rate as an adjustable parameter. They find that an elevated merging rate may completely re-shape the measured SMF (their Figure 9).

While purely mass-independent environmental quenching does not reproduce our results, as discussed in Sect.~\ref{sec:discussion2}, the inclusion of ``early mass quenching'' brings the model predictions much closer to the measured SMFs. Further observables, such as measured ages of stellar populations in different environment, are required to critically test this model (as we will discuss in future work).

In this subsection we have treated the cluster and field environments as completely separate environments, with either different formation times, or different star formation (or gas consumption) rates. This is an obvious limitation, as clusters grow by the accretion of surrounding structures that may all have different formation/collapse times \citep[some being the sites where we expect pre-processing to be taking place, Reeves et al., in prep.][]{mcgee09,fossati17}. We further note that a fundamental limitation of the approach we have taken is that the \citet{peng10} model is not a physical model. The assumption that the quenching rate is proportional to the instantaneous SFR seems difficult to explain in physical terms, even though it helps to reproduce many observational results. In the next section we thus discuss a more physically motivated model to interpret our findings.

\subsection{Results in the light of a ``cosmic starvation'' scenario}\label{sec:discussion1}
Some studies have argued that a quenching mechanism dubbed ``cosmic starvation'' (or ``strangulation'') may be responsible for the majority of quenching observed in the Universe \citep[][]{Larson1980,peng15,fillingham15,davies16,trussler20}, both for quenching that works as a function of mass and environment. In the context of this work, it is assumed that the accretion flow of gas is cut-off once a galaxy has become a satellite of a larger halo \citep[cf.~Fig.~13 in][]{schawinski14}. For instance, one could say that the gas accretion is cut-off once the galaxy is part of a halo with $M_{\mathrm{halo}}\gtrsim 10^{12}\,\mathrm{M_{\odot}}$ \citep{dekel06}. What follows is that the galaxy will consume its left-over gas supply until it is entirely ``starved'' and quenches. 

Outflows associated with star formation are expected to further shorten the gas depletion times compared to that which is expected for ``cosmic starvation''. In such a process, dubbed ``overconsumption'' by \citet{mcgee14}, outflows with typical mass loading factors $\eta \sim 2-3$ would shorten the total quenching/delay time substantially \citep{balogh16}. In this way, galaxies may already quench before stripping events occur \citep{mcgee14}. Because star formation rates were much higher in the past, a key feature of this model is that ``overconsumption'' is very effective at high redshift. Another feature is that high-mass star-forming galaxies quench their star formation relatively more efficiently after cut-off from cosmic gas inflows, since empirical relations show that lower-mass galaxies have relatively larger gas reservoirs and thus longer gas consumption time scales. This is fully in line with what we observe in the shape of the SMF of quiescent galaxies. Also, the general dependence of the $QFE$ on stellar mass is well reproduced by this model \citep{balogh16}.

However, it is not clear whether ``overconsumption'' makes a matching prediction for the measured shape of the SMF of star-forming cluster galaxies.
If massive galaxies are cut-off from their gas supply (either in the present cluster environment, or in their pre-processing stage), they would quench and be removed from the parent population of star-forming galaxies. We would naively expect this to affect the shape of the SMF of star-forming galaxies. Yet, we observe the shape of SMF of star-forming galaxies to be identical between cluster and field, and this seems to be at odds with the predictions of this model. However, we note that the abundance of high-mass galaxies are least well constrained with our data, and there may be enough flexibility in the data to fully match the predictions of ``overconsumption''.

Ultimately, one would utilise cosmological theoretical models of galaxy formation to interpret our measurements. In contrast to the \citet{peng10} framework, hydrodynamical simulations or semi-analytic models may provide insight into the physical processes at play. While the \emph{total} galaxy SMF at $z < 4$ can be reproduced by the current generation of hydrodynamical simulations \citep{furlong15,pillepich18} and semi-analytic models \citep[e.g.][]{henriques15,delucia19}, environment-specific quantities are still challenging to model correctly. By $z \sim 0$, satellite galaxies in simulations are over-quenched in dense environments \citep[e.g.][]{weinmann12}. Possible explanations for this include excessive densities of the intra-cluster medium, an underestimation of a galaxy's ability to hold on to its gas due to finite resolution and/or a lack of a dense, cold ISM component in the simulations \citep{bahe17, kukstas19}. The results presented in this paper provide an additional observational constraint, at higher $z$, to be compared against, with the hope that this may provide additional insights in where the current simulations and theoretical models fail and may be improved (Kukstas et al., in prep.).

\section{Summary and conclusions}\label{sec:summary}
We measure and study the SMF of star-forming and quiescent galaxies in 11 GOGREEN clusters at $1.0<z<1.4$. Thanks to deep multi-band photometry that spans (at least) B/$g$-band to 4.5$\mu$m, we are able to measure the SMF down to stellar masses of $10^{9.7}\,\mathrm{M_{\odot}}$ ($10^{9.5}\,\mathrm{M_{\odot}}$ for star-forming galaxies), which makes this the most precise SMF measured in high-$z$ dense environments. A critical aspect of these measurements is the support by extensive and deep mass-selected spectroscopic sampling with Gemini/GMOS. In particular, this allows us to perform a much cleaner and more precise accounting and removal of fore- and background interlopers (compared to the ordinary statistical subtraction of fore- and background interlopers based on a reference blank field). We compare the cluster galaxy SMF to that measured for the co-eval COSMOS/UltraVISTA field, at similar depth, in order to investigate which processes are responsible for quenching galaxies in different environments. Our main findings are:

\begin{itemize}
\setlength\itemsep{0.3cm}
\item The clusters have a much higher quenched fraction than the co-eval field, over the whole stellar mass range
\item Yet, the SMF of quiescent galaxies has an indistinguishable shape, within the uncertainties of our data, between the GOGREEN cluster and co-eval field.
\item The shape of the SMF of star-forming galaxies is also indistinguishable between cluster and field galaxies.
\item Despite the identical shapes of the SMFs of the two galaxy types, clusters have a different total SMF than the co-eval field environment. This is a reflection of their much-higher quenched fractions than the field.
\item We define the excess quenching due to the cluster environment, on top of a quenching baseline set by the average environment probed by the field survey, as a quenched fraction excess $QFE$. We find $QFE$ to be positive over the entire mass range we probe. This indicates that processes related to the cluster or its formation result in a passive fraction elevated with respect to field galaxies at all stellar masses.
\item The $QFE$ strongly increases with increasing stellar mass, from $\sim$30\% at $M_{\star}=10^{9.7}\,\rm{M_{\odot}}$, to $\sim$80\% at $M_{\star}=10^{11.0}\,\rm{M_{\odot}}$. This is in stark contrast to many studies of the local Universe, in which the $QFE$ in dense environments is found to have no mass dependence at a high level of significance.
\end{itemize}

Whatever process is responsible for the environmental excess quenching at this early epoch, it must therefore be strongly dependent on stellar mass (or masquerade as such at the epoch of observation). We have discussed such options in the context of several galaxy quenching models:
\begin{itemize}
\setlength\itemsep{0.3cm}
\item Pure mass-independent environmental quenching (as in \citet{peng10}) assumes a $QFE$ that is independent of mass. This is opposite to our finding that the $QFE$ is a strongly increasing function of stellar mass.
\item An earlier formation time, compared to the field, of the progenitors of our cluster galaxies would be able to explain our mass-dependent $QFE$ (``mass quenching'' according to \citet{peng10}). However, this scenario predicts stellar population ages of quiescent galaxies that vary in different environments. It is questionable whether this is consistent with our data, as we discuss in Webb et al. (in prep.).
\item We argue that a physically motivated model of ``overconsumption'', as introduced by \citet{mcgee14}, may provide an explanation of the observed trends. As discussed, such a model is expected to leave an imprint at the high-mass end of the SMF of star-forming cluster galaxies. It is still unclear whether this is consistent with the data.
\end{itemize}

Our results unambiguously point at a quenching mechanism that works differently from what we observe in the local Universe. It is likely that the current set of models provide the general framework for galaxy quenching in different environments, also at higher redshift. However, now that more precise measurements are available in high-$z$ over-dense environments, they would have to be revised to provide also an accurate representation of the distant universe.

\begin{acknowledgements}
We thank the anonymous reviewer for useful comments that helped to improve and clarify the message of this work. It is a pleasure to thank Ian McCarthy and Allison Man for insightful discussions. We thank Sean Fillingham, Callum Bellhouse, Melinda Townsend and Nicole Drakos for help with the observations. RvdB thanks George Lansbury and Marianne Heida for kindly providing a computer screen that allowed him to efficiently work remotely.
We thank the International Space Science Institute (ISSI) for providing financial support and a meeting facility that inspired insightful discussions for team ``COSWEB: The Cosmic Web and Galaxy Evolution''.

G.R. acknowledges support from the National Science Foundation grants AST-1517815, AST-1716690, and AST-1814159 and NASA HST grant AR-14310.  GR also acknowledges the support of an ESO visiting science fellowship.
R.D. gratefully acknowledges support from the Chilean Centro de Excelencia en Astrof\'isica y Tecnolog\'ias Afines (CATA) BASAL grant AFB-170002. The European Space Agency (ESA) Research Fellowship (LJO). Universidad Andr\'es Bello Internal Project \#DI-12-19/R (JN). P.C. acknowledges the support of the ALMA-CONICYT grant no 31180051.

This work is supported by the National Science Foundation through grant AST-1517863, by \textit{HST} program number GO-15294, and by grant number 80NSSC17K0019 issued through the NASA Astrophysics Data Analysis Program (ADAP). Support for program number GO-15294 was provided by NASA through a grant from the Space Telescope Science Institute, which is operated by the Association of Universities for Research in Astronomy, In- corporated, under NASA contract NAS5-26555.

This work was supported in part by NSF grants AST-1815475 and AST-1518257. Additional support was provided by NASA through grant AR-14289 from the Space Telescope Science Institute, which is operated by the Association of Universities for Research in Astronomy, Inc., under NASA contract NAS 5-26555. 

This project has received funding from the European Research Council (ERC) under the European Union’s Horizon 2020 research and innovation programme (grant agreement No 769130). 
This project has received funding from the European Research Council (ERC) under the European Union's Horizon 2020 research and innovation programme (grant agreement No. 833824). BV acknowledges financial contribution from the grant PRIN MIUR 2017 n.20173ML3WW\_001 (PI Cimatti) and from the INAF main-stream funding programme (PI Vulcani). 

Based on observations obtained at the Gemini Observatory (GS LP-1 and GN LP-4), which is operated by the Association of Universities for Research in Astronomy, Inc., under a cooperative agreement with the NSF on behalf of the Gemini partnership: the National Science Foundation (United States), National Research Council (Canada), CONICYT (Chile), Ministerio de Ciencia, Tecnolog\'{i}a e Innovaci\'{o}n Productiva (Argentina), Minist\'{e}rio da Ci\^{e}ncia, Tecnologia e Inova\c{c}\~{a}o (Brazil), and Korea Astronomy and Space Science Institute (Republic of Korea).

Based on observations collected at the European Organisation for Astronomical Research in the Southern Hemisphere under ESO programmes 097.A-0734(A) and 097.A-0734(B).

Based on observations obtained with MegaPrime/MegaCam, a joint project of CFHT and CEA/DAPNIA, at the Canada-France-Hawaii Telescope (CFHT) which is operated by the National Research Council (NRC) of Canada, the Institut National des Sciences de l'Univers of the Centre National de la Recherche Scientifique of France, and the University of Hawaii. 

Based on observations obtained with WIRCam, a joint project of CFHT, the Academia Sinica Institute of Astronomy and Astrophysics (ASIAA) in Taiwan, the Korea Astronomy and Space Science Institute (KASI) in Korea, Canada, France, and the Canada-France-Hawaii Telescope (CFHT) which is operated by the National Research Council (NRC) of Canada, the Institut National des Sciences de l'Univers of the Centre National de la Recherche Scientifique of France, and the University of Hawaii. 

This paper includes data gathered with the 6.5 meter Magellan Telescopes located at Las Campanas Observatory, Chile.

This work is based in part on observations made with the \textit{Spitzer Space Telescope}, which is operated by the Jet Propulsion Laboratory, California Institute of Technology under a contract with NASA.

Based in part on data collected at Subaru Telescope, which is operated by the National Astronomical Observatory of Japan.

The Hyper Suprime-Cam (HSC) collaboration includes the astronomical communities of Japan and Taiwan, and Princeton University. The HSC instrumentation and software were developed by the National Astronomical Observatory of Japan (NAOJ), the Kavli Institute for the Physics and Mathematics of the Universe (Kavli IPMU), the University of Tokyo, the High Energy Accelerator Research Organization (KEK), the Academia Sinica Institute for Astronomy and Astrophysics in Taiwan (ASIAA), and Princeton University. Funding was contributed by the FIRST program from Japanese Cabinet Office, the Ministry of Education, Culture, Sports, Science and Technology (MEXT), the Japan Society for the Promotion of Science (JSPS), Japan Science and Technology Agency (JST), the Toray Science Foundation, NAOJ, Kavli IPMU, KEK, ASIAA, and Princeton University.  

This research has made use of the SVO Filter Profile Service (http://svo2.cab.inta-csic.es/theory/fps/) supported by the Spanish MINECO through grant AYA2017-84089.
\end{acknowledgements}

\bibliographystyle{aa} 
\bibliography{MasterRefs} 

\begin{appendix}

\section{Robustness tests}
In this Appendix we address several assumptions we had to make in this analysis, and study their impact on the presented results.

\subsection{Magnitude bias and Eddington bias}\label{app:eddington}
\begin{figure}
\resizebox{\hsize}{!}{\includegraphics{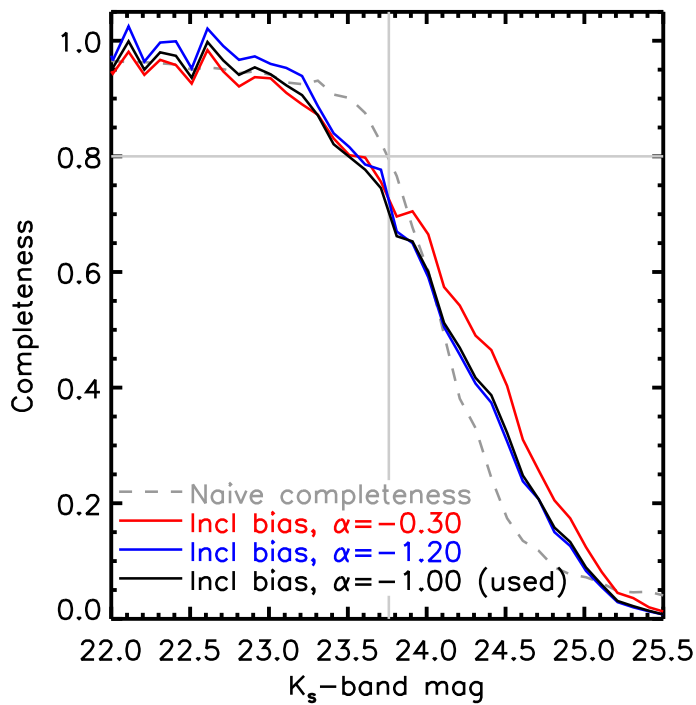}}
\caption{Completeness of the detected sources as a function of magnitude. This is for a cluster at median $K_\mathrm{s}$-band depth. \textit{Grey dashed:} Naive correction factor in the absence of magnitude bias and Eddington bias. Here the x-axis refers to intrinsic magnitudes, and the y-axis to the fraction of these sources that is recovered (at any measured magnitude). \textit{Solid lines:} Completeness as a function of recovered magnitude for different intrinsic magnitude distributions (indicated by different colours). Here 1/Completeness would be the factor by which one needs to multiply the measured counts at different magnitudes, to recover the true intrinsic magnitude distribution. This corrects implicitly for magnitude- and Eddington bias. While we use the black solid curve in this work, we note that they are very similar in the range where the galaxy population is studied (i.e.~leftward of the vertical grey line).}
\label{fig:Eddingtonbias}
\end{figure}

We conservatively study the galaxy population down to our 80\% detection limit, and perform a small incompleteness correction which is in principle a maximum of 25\% increase in source counts. However, many studies determine those correction factors based on the recovery percentage of \textit{injected} sources, and use these to correct counts at \textit{measured} magnitudes (cf.~Fig.~\ref{fig:Eddingtonbias}). One should however perform this correction to the distribution as a function of measured magnitudes, and there is generally a bias between the intrinsic and measured magnitudes. Especially around the detection limit, sources are systematically measured to be fainter (by about 0.1 magnitudes) than their intrinsic magnitudes (at least the AUTO/Kron-like magnitudes measured with our \texttt{SExtractor} setup, as determined from the image simulations).

In combination with measurement uncertainties (scatter), this introduces a dependence on the intrinsic magnitude distribution of sources, which results in \citet{eddington1913} bias. For example, for an intrinsically steep magnitude distribution, there will be relatively more faint sources scattering to brighter magnitudes than bright sources scattering to fainter magnitudes. This effect is relatively minor in our case (compare the different solid lines in Fig.~\ref{fig:Eddingtonbias}). Noting that our SMFs are measured in logarithmic units, even the general magnitude bias does not have a large impact on our results (compare the solid lines with the dashed line in Fig.~\ref{fig:Eddingtonbias}). Properly accounting for Eddington bias would require to perform the measurement iteratively \citep[as in][]{vanderburg10}, but for this work we will assume a faint-end slope with power-law slope $\alpha = -1$, which is somewhere between the ones we measure for quiescent and star-forming galaxies, in estimating the Eddington bias.

\subsection{Spectroscopic target selection}\label{app:spectargets}
\begin{figure*}
\resizebox{\hsize}{!}{\includegraphics{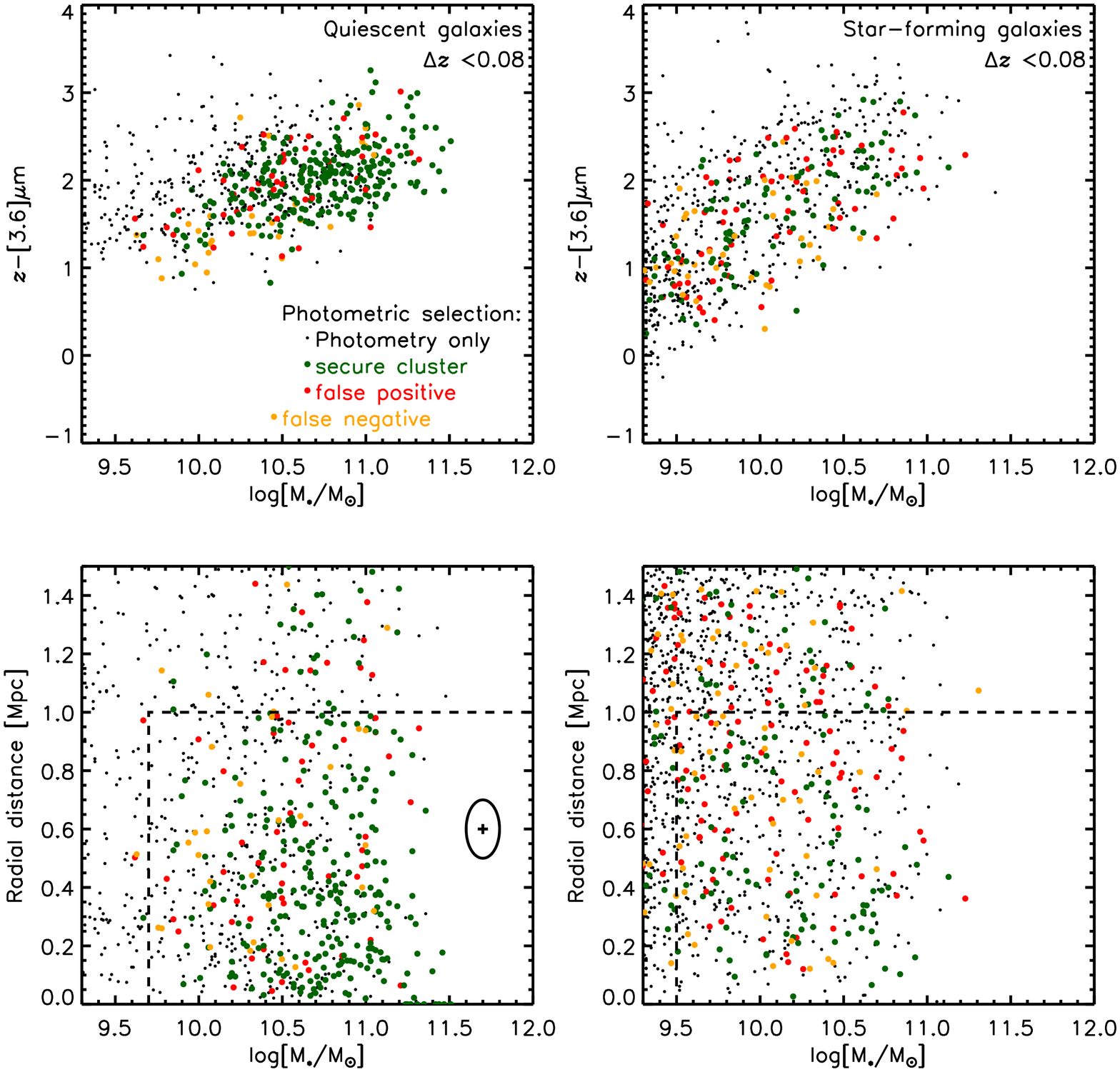}}
\caption{\textit{Top panels:} Colour- stellar mass diagram that compares the spectroscopic targets (coloured points) with photometrically selected cluster galaxy candidates (black+green+red). Our method of subtracting fore- and background interlopers relies on the spectroscopic subset being a representative subset. \textit{Left panel:} Quiescent galaxies, for which this criterion is not fully met at low masses. \textit{Right panel:} Star-forming galaxies, for which this assumption is valid. \textit{Lower panels:} Spectroscopic targets shown as a function of cluster-centric distance and stellar mass. The ellipse shows the relative weight we give to either parameter when defining the five closest sources in this plane (which are used to estimate membership for the black points). Dashed region defines the population studied in this work.}
\label{fig:spectargets}
\end{figure*}
Our fiducial method to measure the cluster galaxy SMF relies on the assumption that the spectroscopic sample is representative of the total galaxy population. We test this assumption in Fig.~\ref{fig:spectargets}, where we compare, in the upper two panels, the photometric sample with the spectroscopic targets in a plane of colour and stellar mass. The black points, of which the green and red are subsets, are the sources selected to be cluster galaxy candidates based on their photometric redshifts. The coloured points trace this distribution well for the star-forming population (right panel), whereas they do not entirely trace the photometrically selected distribution for the quiescent population (left panel), especially at the lowest masses. 

There are several reasons for this mis-match. Firstly, the photometric redshifts were not used to select spectroscopic targets. Rather, the selection was based on $z-\mathrm{[3.6]}$ colour versus $z$ magnitude \citep{balogh17}. Since quiescent and star-forming galaxies have different $M_{\star}/L$, this does not map in a straightforward way to the stellar-mass dimension used in this work. Secondly, red quiescent galaxies with low stellar masses are extremely faint in the blue, making the estimation of absorption-line based redshifts very challenging, even with 15-hour long integrations. For this reason we perform an additional robustness test in Sect.~\ref{app:approach2}.

The lower two panels of Fig.~\ref{fig:spectargets} illustrate the spectroscopic target selection as a function of stellar mass and cluster-centric radius. In this plane we select, as described in Sect.~\ref{sec:membership}, for each un-targeted galaxy (black points), five similar galaxies that \textit{were} targeted. To find the five closest galaxies, in stellar mass and distance, a 0.1 \texttt{dex} difference in stellar mass is taken to be equivalent to a 100 kpc difference in radial distance (illustrated by the ellipse in the lower left panel).

Even though the spectroscopic sampling is not complete, we have spectroscopic measurements over the entire range in stellar mass and radial distance. The dashed lines illustrate the boundaries within which we study the galaxy populations in the GOGREEN clusters, but we use spectroscopic targets slightly outside the box to perform the membership correction.

\subsection{Statistical background subtraction}\label{app:approach2}
While for the star-forming cluster galaxies we find that the spectroscopic sub-sample is well representative of the full photometric sample (Appendix~\ref{app:spectargets}), this is not entirely true for the quiescent population, especially at the faint/low-mass end of the distribution (as explained in Appendix~\ref{app:spectargets}). To investigate if this may have caused a bias in the measured low-mass end of the quiescent SMF, we perform an analysis that does not rely on the spectroscopic data set at all. Since the quiescent targets are reasonably over-dense against fore- and background interlopers, we subtract these interlopers statistically. We have made use of the COSMOS/UltraVISTA survey \citep{muzzin13a}, and performed an analysis using only a sub-set of the available filters to make it more representative of the GOGREEN photometric data set. 

\begin{figure}
\resizebox{\hsize}{!}{\includegraphics{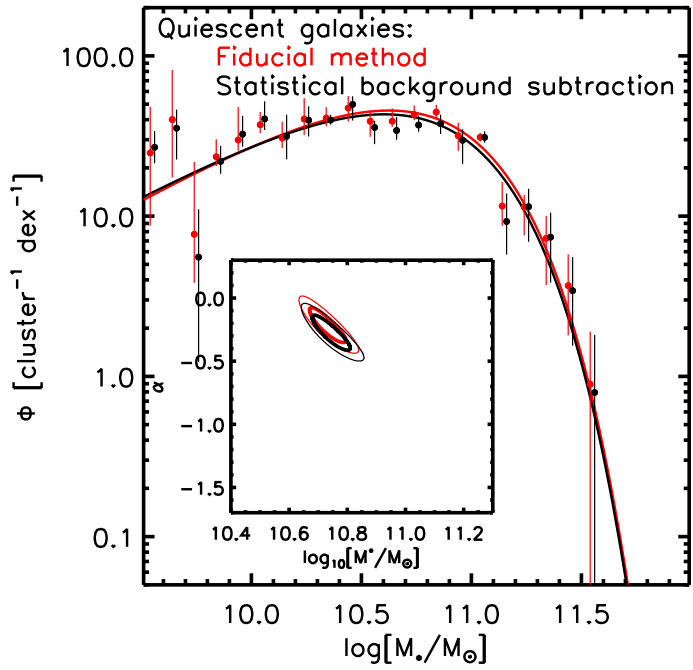}}
\caption{\textit{Red:} Our fiducial analysis makes best use of the spectroscopic coverage of GOGREEN. It reduces the effect of cosmic variance since it does not rely on the statistical subtraction of interlopers from a reference field. However, for this approach we have to assume that the spectroscopic subset is representative of the full galaxy population. \textit{Black:} An alternative approach is to subtract the fore- and background interlopers statistically by making use of the external COSMOS/UltraVISTA survey. The latter approach does not make use of the spectroscopy at all, yet presents results that are fully consistent with those obtained with the fiducial method. Small horizontal offsets have been applied, to the data points only, for better visibility.}
\label{fig:m1m2comparison}
\end{figure}

The comparison is made in Fig.~\ref{fig:m1m2comparison}, showing good consistency, within the uncertainties, of the results based on both methods. We note that, for star-forming galaxies the over-density is so low that such an approach would lead to a very imprecise measurement.

For Approach 2, which relies on a statistical background subtraction, such interloper galaxies are included in the same likelihood and have a negative weight. 

\section{Colour images}\label{app:colourimages}
This Appendix presents colour images of the 11 clusters studied here. The physical scale (corresponding to an on-sky angle of 1 arcmin) is indicated for each cluster.

\begin{figure*}
\centering
\begin{minipage}{.495\textwidth}
  \centering
  \includegraphics[width=.90\linewidth]{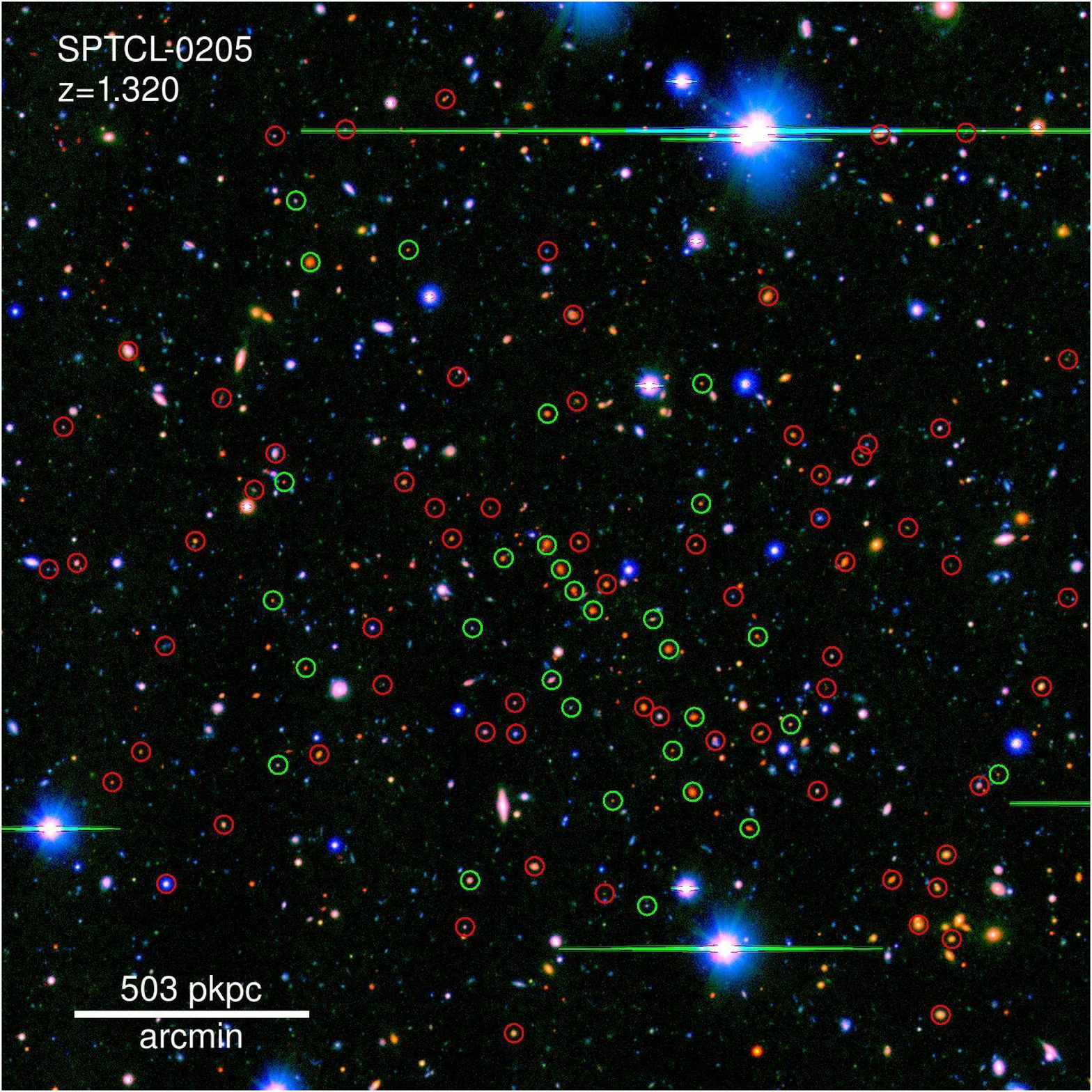}
\end{minipage}
\begin{minipage}{.495\textwidth}
  \centering
  \includegraphics[width=.90\linewidth]{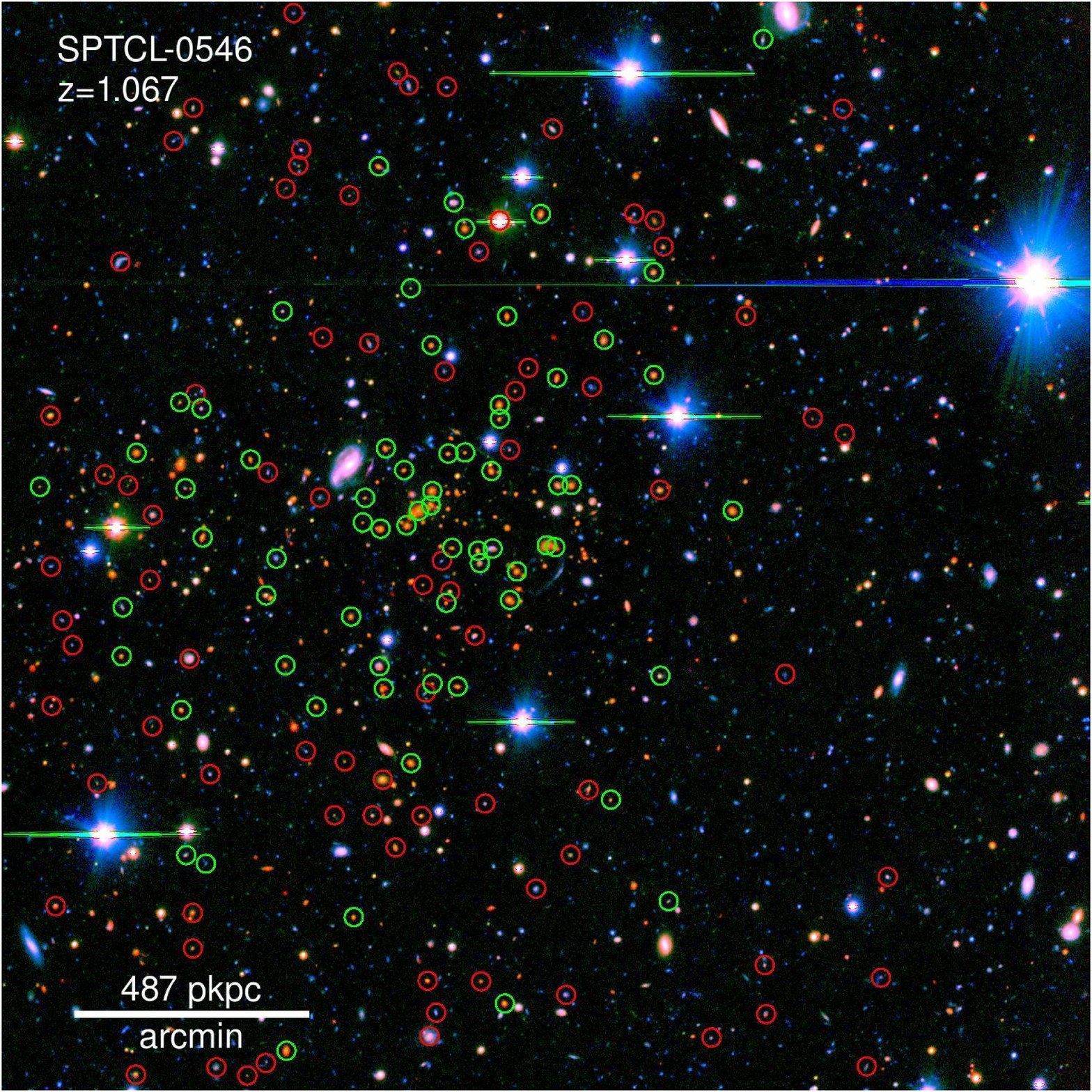}
\end{minipage}
\begin{minipage}{.495\textwidth}
  \centering
  \includegraphics[width=.90\linewidth]{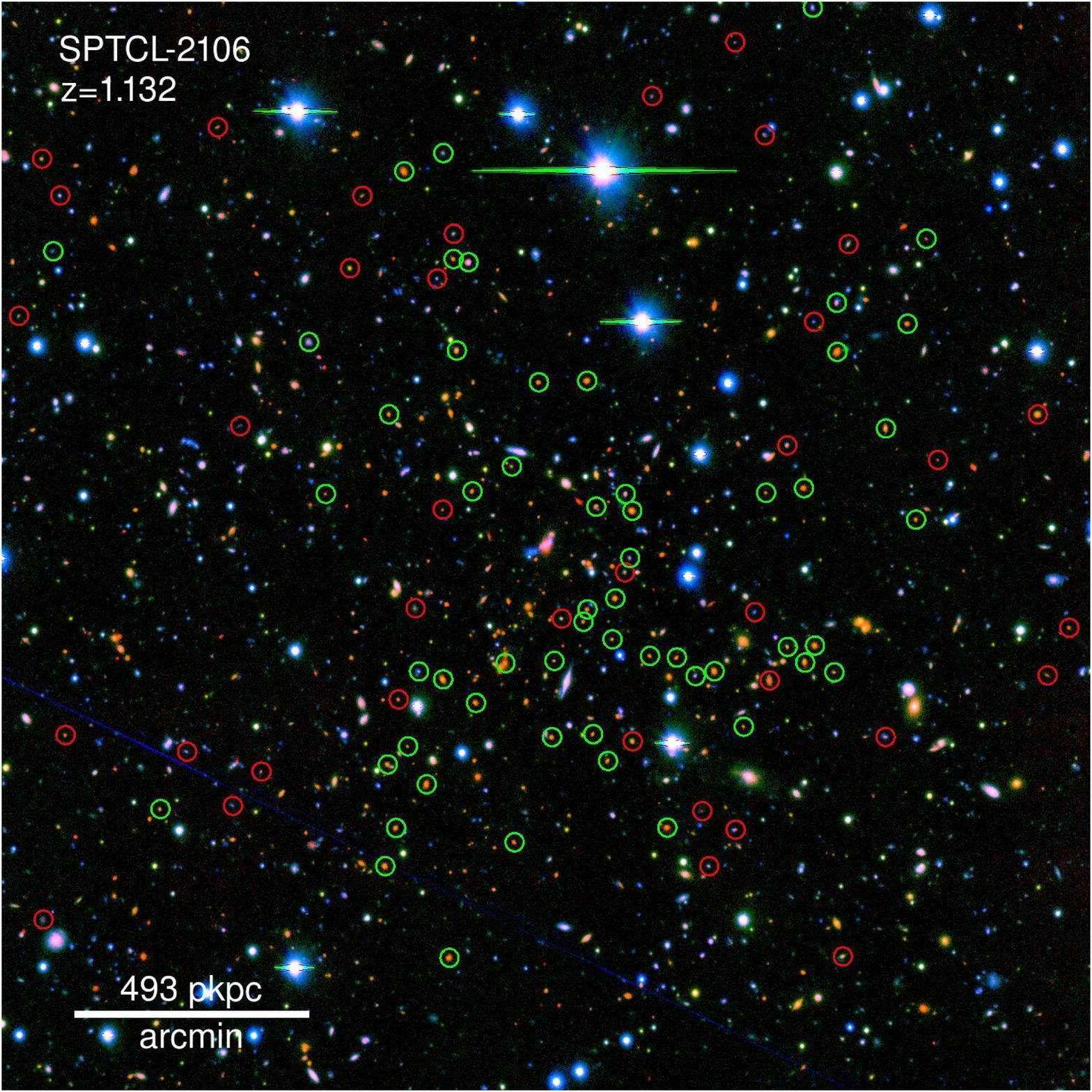}
\end{minipage}
\begin{minipage}{.495\textwidth}
  \centering
  \includegraphics[width=.90\linewidth]{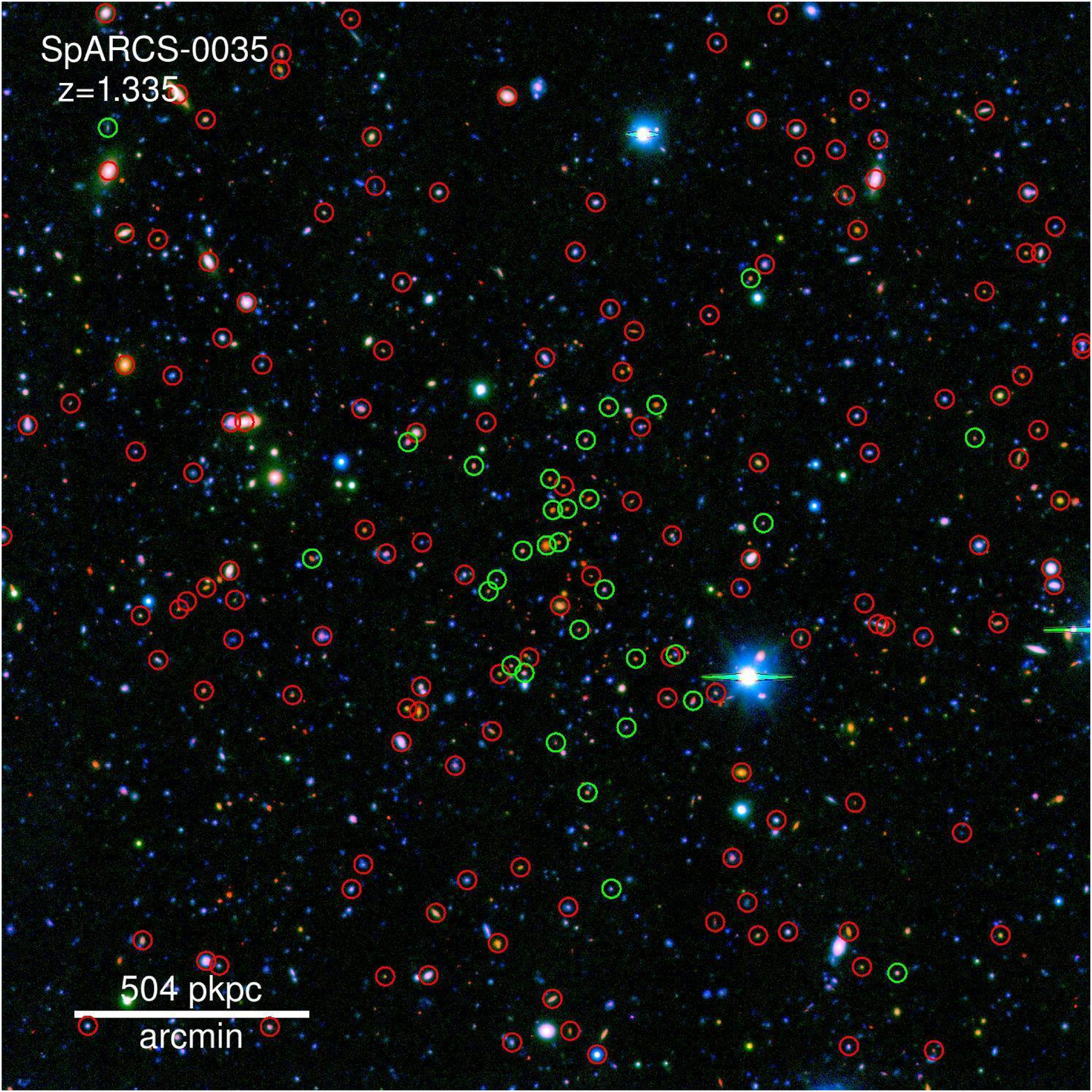}
\end{minipage}
\begin{minipage}{.495\textwidth}
  \centering
  \includegraphics[width=.90\linewidth]{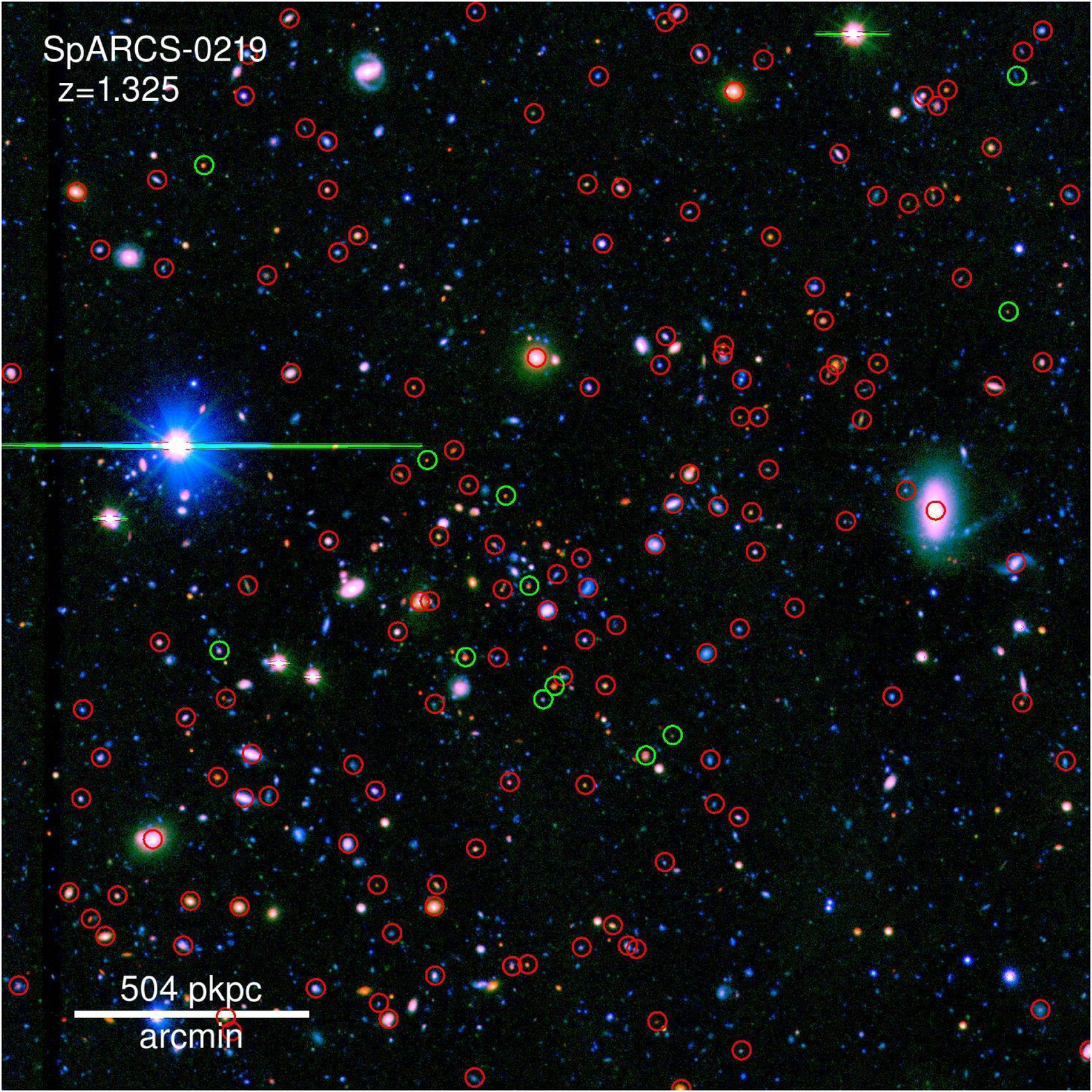}
\end{minipage}
\begin{minipage}{.495\textwidth}
  \centering
  \includegraphics[width=.90\linewidth]{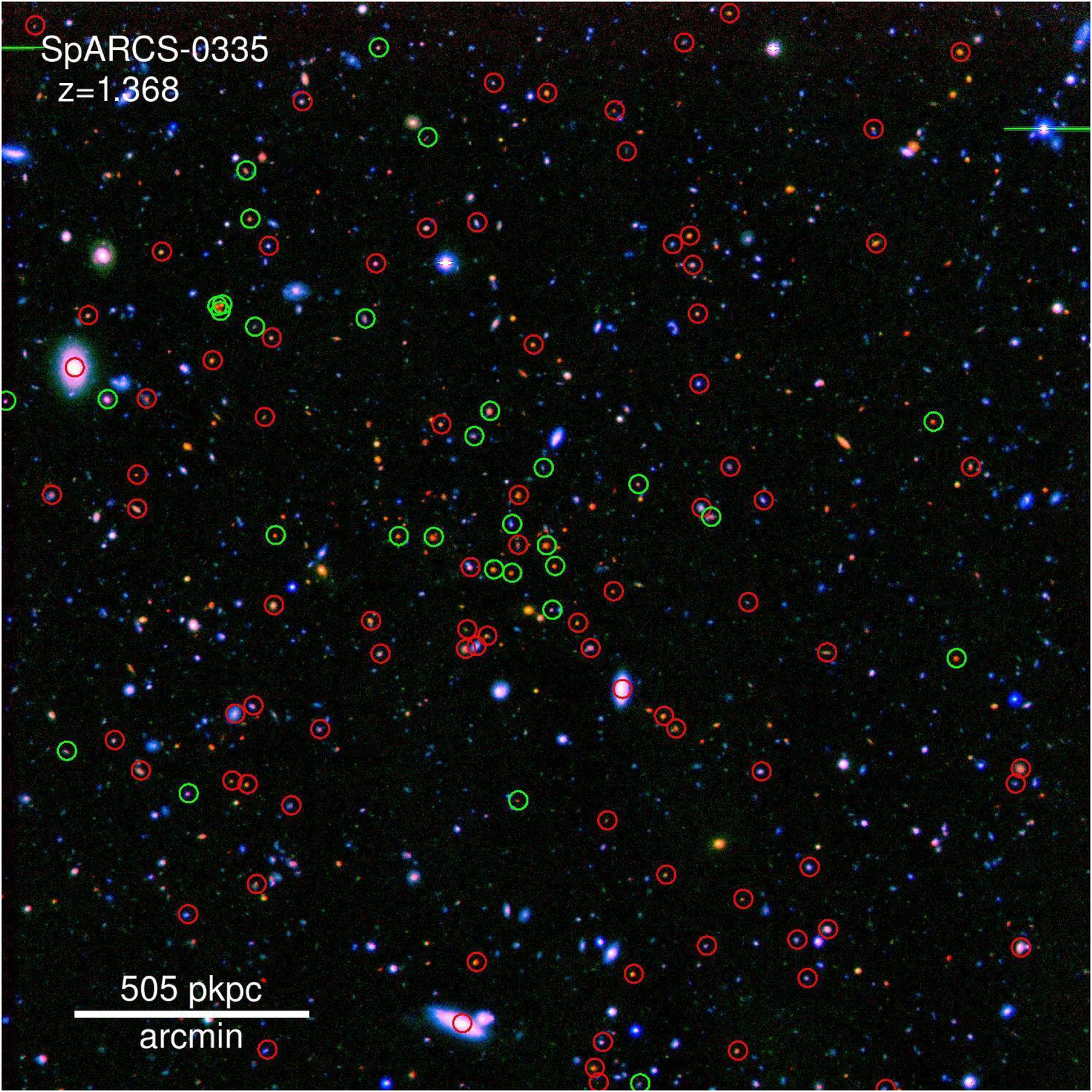}
\end{minipage}
\caption{Colour composite images of the clusters in our sample, based on $g$ or $B$-, $i/I$-, and $K_\mathrm{s}$-band imaging. Spectroscopic targets are indicated, with cluster members in green and non-members in red. An angular scale is indicated, together with the physical scale transverse to the line-of-sight.}
\label{fig:gallery1}
\end{figure*}
\begin{figure*}
\centering
\begin{minipage}{.495\textwidth}
  \centering
  \includegraphics[width=.90\linewidth]{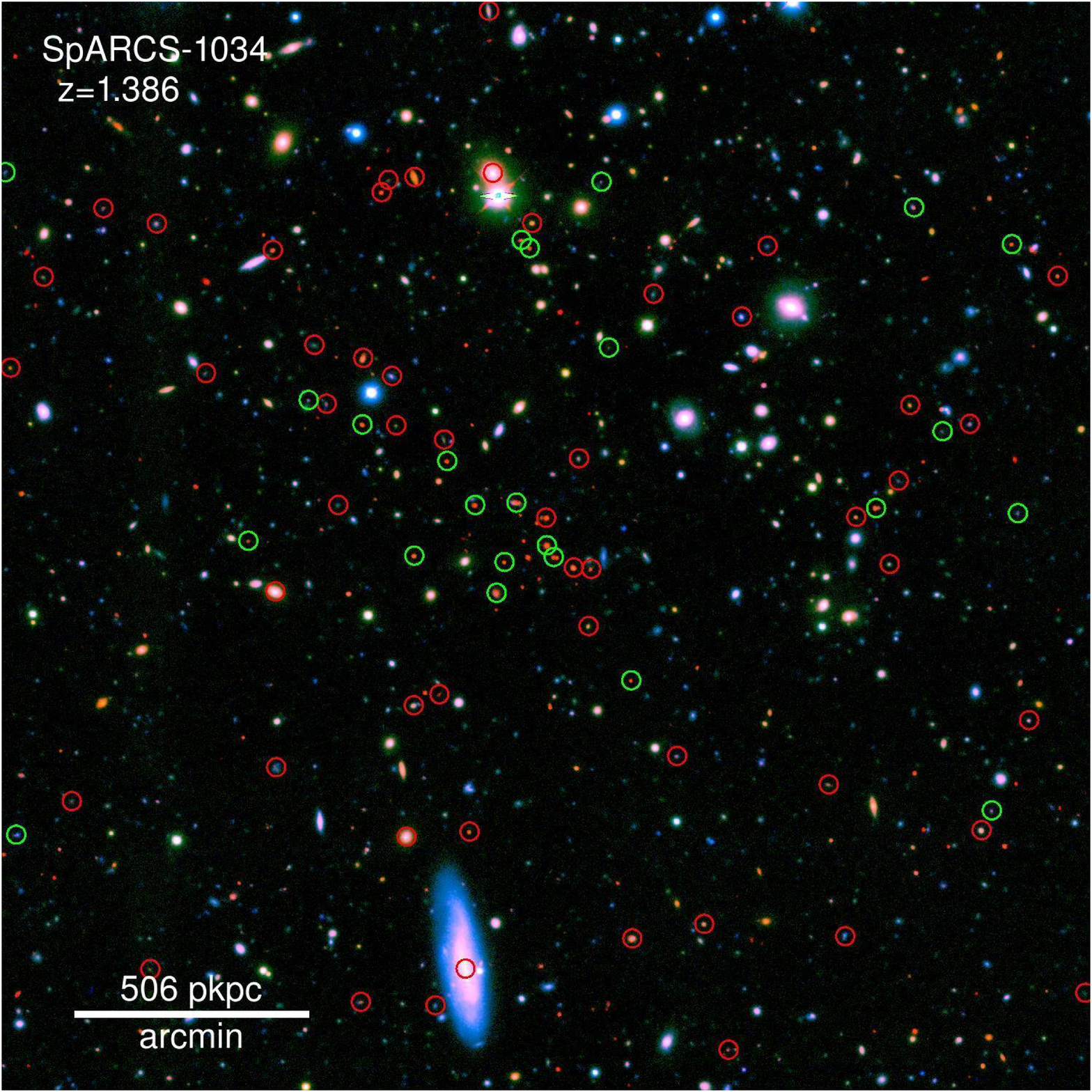}
\end{minipage}
\begin{minipage}{.495\textwidth}
  \centering
  \includegraphics[width=.90\linewidth]{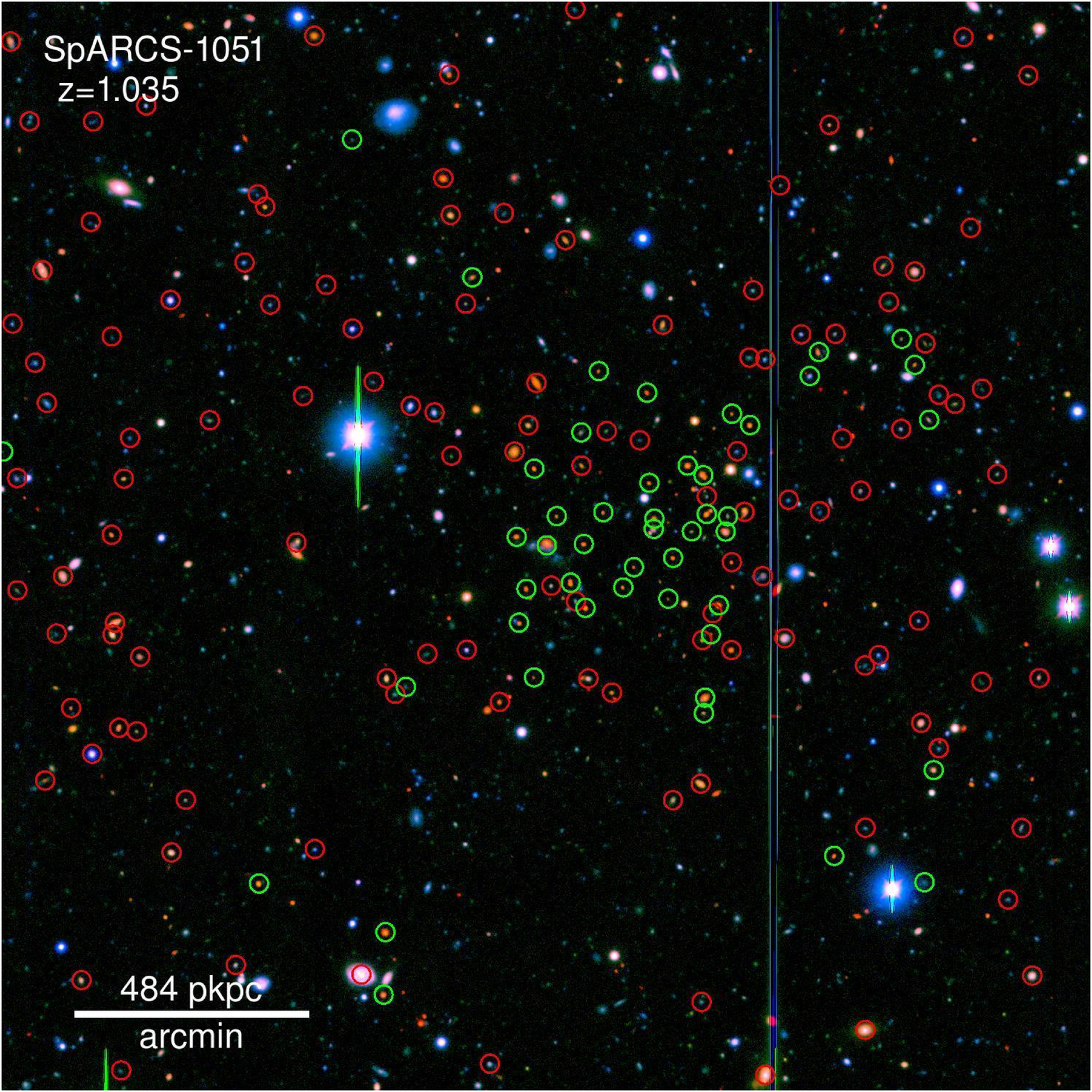}
\end{minipage}
\begin{minipage}{.495\textwidth}
  \centering
  \includegraphics[width=.90\linewidth]{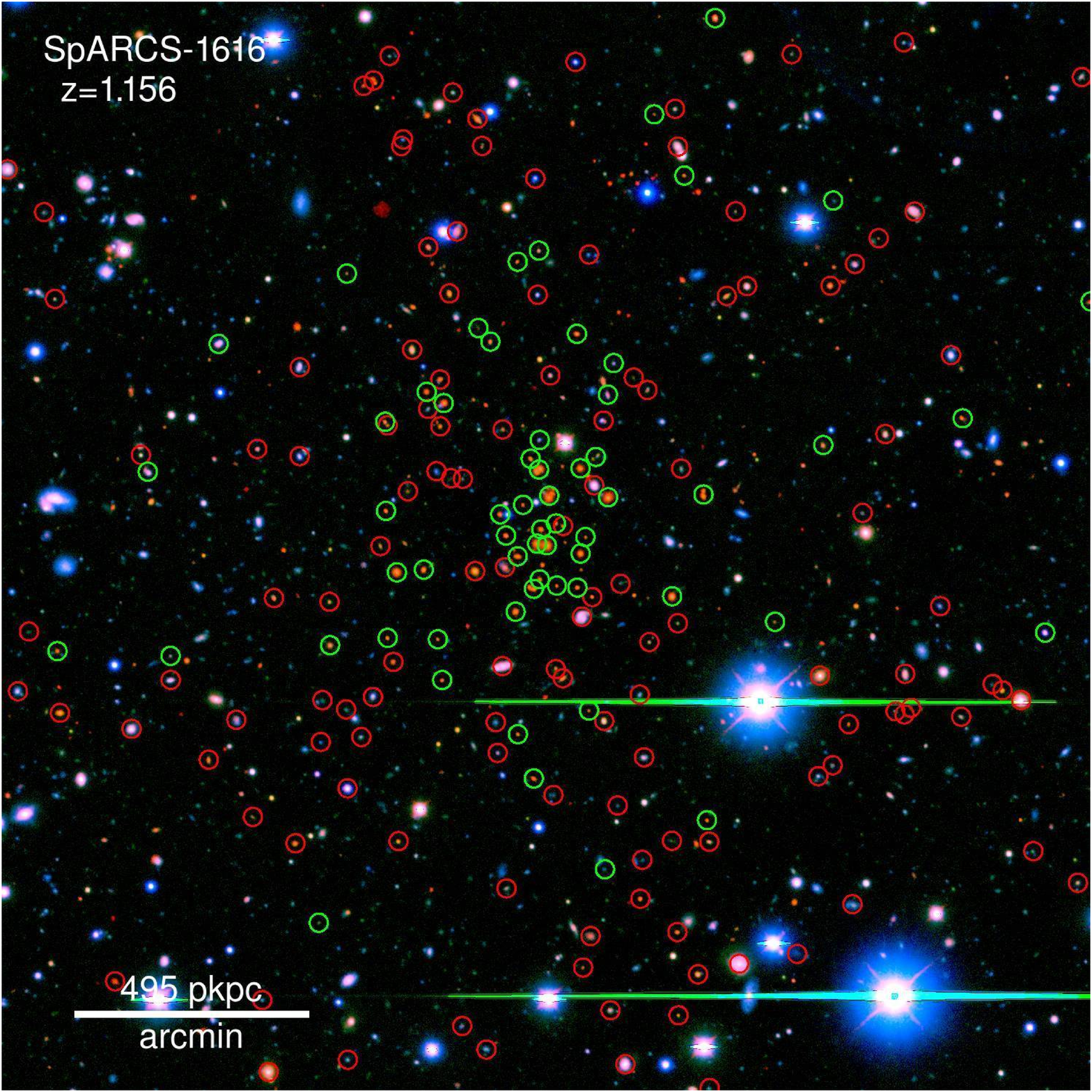}
\end{minipage}
\begin{minipage}{.495\textwidth}
  \centering
  \includegraphics[width=.90\linewidth]{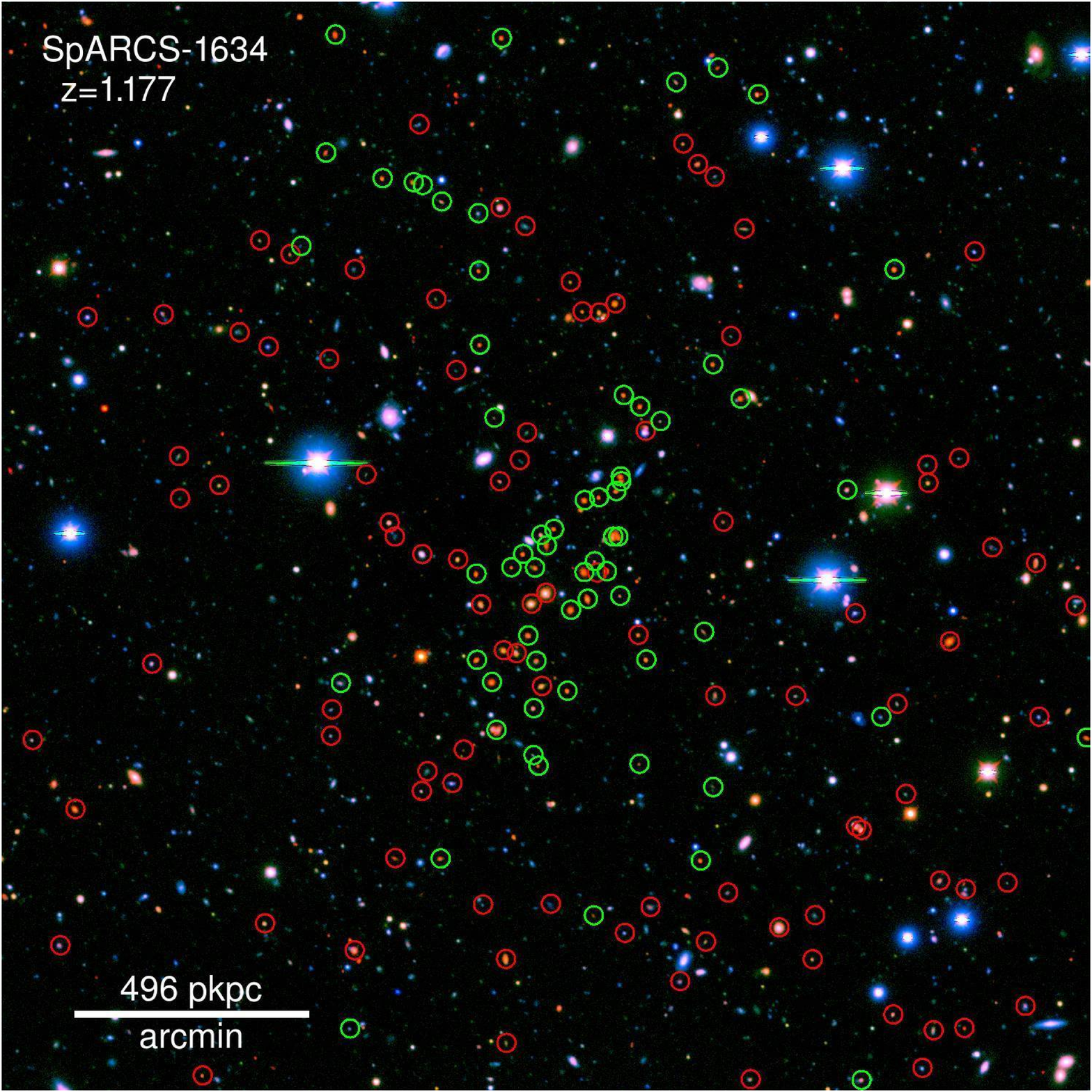}
\end{minipage}
\begin{minipage}{.495\textwidth}
  \centering
  \includegraphics[width=.90\linewidth]{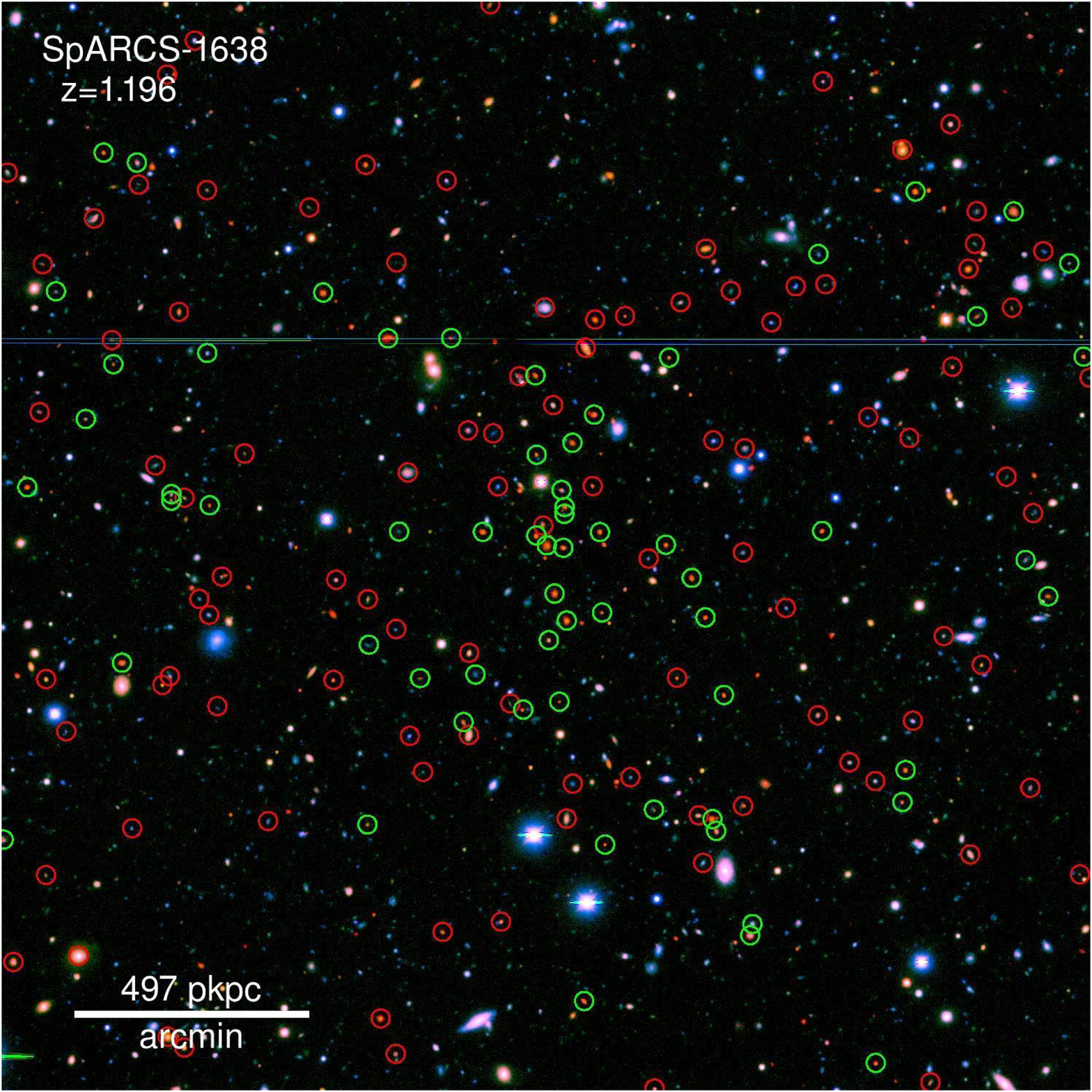}
\end{minipage}
\caption{... continued.}
\label{fig:gallery2}
\end{figure*}

\end{appendix}

\end{document}